\documentclass[11pt,a4paper]{article}
\pdfoutput=1
\usepackage{amsmath}
\usepackage{amsfonts}
\usepackage{color}
\usepackage{graphicx}% Include figure files
\usepackage{dcolumn}% Align table columns on decimal point
\usepackage{bm}
\usepackage[%dvipdfm,
bookmarks,bookmarksnumbered,colorlinks=true,anchorcolor=blue,
linkcolor=blue,urlcolor=blue,citecolor=blue,%hypertex,
breaklinks=true]{hyperref}
\usepackage{geometry}
\usepackage{authblk}
\usepackage{amssymb}
\usepackage[utf8]{inputenc}
\usepackage[numbers,sort&compress]{natbib}
\usepackage{ulem}

\numberwithin{equation}{section}   % numerate equations by section

\usepackage{xcolor}

\date{\today}

\begin{document}
%\title{\bf Fluctuating holographic superconductor}
%\title{\bf Fluctuating order parameter in a holographic superconductor}
\title{\bf Ginzburg-Landau effective action for a fluctuating holographic superconductor}

\author[1]{Yanyan Bu \thanks{yybu@hit.edu.cn}}
\author[2]{Mitsutoshi Fujita \thanks{fujita@mail.sysu.edu.cn (co-correspondence author)}}
\author[2,3]{Shu Lin \thanks{linshu8@mail.sysu.edu.cn (co-correspondence author)}}

\affil[1]{\it School of Physics, Harbin Institute of Technology, Harbin 150001, China} %\authorcr \it
\affil[2]{\it School of Physics and Astronomy, Sun Yat-Sen University, Zhuhai 519082, China }
\affil[3]{\it Guangdong Provincial Key Laboratory of Quantum Metrology and Sensing, Sun Yat-Sen University, Zhuhai 519082, China}
%\authorcr \it

\maketitle

\begin{abstract}
Under holographic prescription for Schwinger-Keldysh closed time contour for non-equilibrium system, we consider fluctuation effect of the order parameter in a holographic superconductor model. Near the critical point, we derive the time-dependent Ginzburg-Landau effective action governing dynamics of the fluctuating order parameter. In a semi-analytical approach, the time-dependent Ginzburg-Landau action is computed up to quartic order of the fluctuating order parameter, and first order in time derivative.

\end{abstract}

\newpage

%\tableofcontents

\allowdisplaybreaks

\flushbottom

\section{Introduction}

The phenomena of phase transition are ubiquitous in nature. A paradigm for describing a wide range of physical systems near the critical point is the phenomenological Ginzburg-Landau theory \cite{Landau-Lifshitz,Kardar2007}. Within this approach, an order parameter field $\Delta(\vec x)$ is associated with the physical system, and a Ginzburg-Landau Hamiltonian density $\mathcal H_{\rm GL}$ is proposed based on symmetry consideration:
\begin{align}
\mathcal H_{\rm GL}[\Delta] = a |\Delta|^2 + b |\Delta|^4+ c(\vec \nabla \Delta)^2 + \cdots + \phi_0 \Delta, \label{H_GL}
\end{align}
where $\phi_0$ is external source for the order parameter, and the $\cdots$ denote higher powers of $\Delta$ as well as those terms including spatial derivatives of $\Delta$. In principle, $\mathcal H_{\rm GL}$ would be obtained by integrating out microscopic degrees of freedom of the system. The phenomenological coefficients $a, b, c$, etc., reflect non-universal microscopic details of the physical system.

The equilibrium partition function is defined as the following functional integral:
\begin{align}
Z_{\rm GL}[\phi_0] = \int D \Delta \, e^{- \beta\int d^3x \mathcal H_{\rm GL}[\Delta] }, \label{Z_GL}
\end{align}
which is usually challenging to compute. In the saddle point (or mean field) approximation, the computation of $Z_{\rm GL}$ is reduced to minimizing $\mathcal H_{\rm GL}$ with respect to variation of $\Delta$. Indeed, the Landau free energy density $\mathcal F_{\rm GL}$ is the spatially uniform limit of $\mathcal H_{\rm GL}$:
\begin{align}
\mathcal F_{\rm GL} = a |\Delta|^2 + b |\Delta|^4 + \phi_0 \Delta. \label{F_GL}
\end{align}

It is desirable to extend the Ginzburg-Landau theory valid for system in equilibrium into a more general framework based on functional integral on closed time contour. The latter is also called Schwinger-Keldysh (SK) formalism \cite{Chou:1984es,Kamenev2011}, which makes the descriptions of quantum systems in and out of equilibrium unified, and becomes an ideal framework for studying real-time dynamics. Within the SK formalism, the quantum system effectively evolves forward (from initial time $t_i$ to final time $t_f$) and then backward (from $t_f$ to $t_i$), forming a closed time contour. Moreover, the SK formalism systematically incorporates both fluctuations and dissipations. This is achieved by the doubling of degrees of freedom: $\Delta \to (\Delta_1, \Delta_2)$, where the subscripts $1,2$ denote the upper and lower branches of the SK closed time contour.

%
%Here, the $(r,a)$-basis is defined as
%\begin{align}
%\Delta_r = \frac{1}{2}(\Delta_1 + \Delta_2), \qquad \Delta_a= \Delta_1 - \Delta_2.
%\end{align}

In recent years, the SK formalism was used to formulate an effective field theory (EFT) for dissipative hydrodynamics \cite{Endlich:2012vt,Kovtun:2014hpa,Nicolis:2013lma,Harder:2015nxa,Grozdanov:2013dba,Crossley:2015evo,Glorioso:2017fpd,Glorioso:2018wxw,Haehl:2015foa,Haehl:2015uoc,Haehl:2018lcu,Jensen:2017kzi,Baggioli:2020haa}. In such an EFT, the dynamical variables are identified with conserved quantities such as energy, momentum or charge density \cite{Crossley:2015evo,Glorioso:2018wxw}. Near the critical point of a phase transition, the order parameter experiences critical slow-down \cite{Glorioso:2018wxw} and becomes an additional slow mode, which should be retained as dynamical variable in such an EFT formulation.

In \cite{Levchenko2007}, Levchenko and Kamenev employed the SK formalism and derived Ginzburg-Landau effective action by systematically integrating out electronic degrees of freedom in the Keldysh nonlinear $\sigma$-model for disordered superconductors. The derivation was carried out in the high temperature phase. The resulting effective action is a functional of the complex order parameter $\Delta$ and external gauge potential $\mathcal A_\mu$. While the derivation of \cite{Levchenko2007} assumes hydrodynamic limit and weak external gauge fields, the effective action goes beyond linear response regime and contains fruitful nonlinear effects. For example, the charge diffusion constant receives a nonlocal correction due to the fluctuation of the order parameter; there is interaction between the fluctuating order parameter $\Delta$ and external field $\mathcal A_\mu$. Schematically, the effective action can be split into three parts: the time-dependent Ginzburg-Landau effective action $S_{\rm GL}$, which is the real-time generalization of $\mathcal H_{\rm GL}$ \eqref{H_GL}; the normal current part $S_{\rm N}$, which describes the dynamics of the charge diffusion; and the supercurrent part $S_{\rm S}$, which is responsible for the interaction between the order parameter and the external gauge field.

This work aims at deriving the Ginzburg-Landau effective action from a holographic perspective \cite{Maldacena:1997re,Gubser:1998bc,Witten:1998qj}, which provides a tractable framework for studying dynamics of strongly coupled large $N_c$ (the number of colors) gauge theory via weakly coupled gravitational theory in asymptotic AdS space. Specifically, we consider a holographic superconductor model \cite{Hartnoll:2008vx,Herzog:2010vz}, in which spontaneously breaking of boundary $U(1)$ symmetry is realized as formation of scalar hair outside the event horizon of Schwarzschild-AdS black hole \cite{Gubser:2008px}. Over the past decade or so, the holographic superconductor model has been examined in various aspects, see review papers or textbooks \cite{Herzog:2009xv,Horowitz:2010gk,Musso:2014efa,Ammon:2015wua,
Zaanen:2015oix,Cai:2015cya,Hartnoll-book2018,Cai:2012sk} and references therein. Interestingly, the Ginzburg-Landau free energy density \eqref{F_GL} was derived in \cite{Yin:2013fwa,Dector:2013dia}\footnote{A bulk Chern-Simons term was found to modify the usual Ginzburg-Landau theory \cite{Banerjee:2013maa}.}, confirming the phase transition is of second order nature; the spectrum of the Goldstone mode associated with spontaneously breaking of $U(1)$ symmetry was identified in \cite{Esposito:2016ria} through bulk fluctuation analysis in pure AdS.

%
%{\color{red} from thermodynamic properties \cite{???} to hydrodynamic transports \cite{Amado:2009ts,Erdmenger:2010xm,Erdmenger:2011tj,Herzog:2011ec,Bhattacharya:2011eea} to far-from-equilibrium time-evolution \cite{Murata:2010dx,Liu,Zhang,Zhang,Tian,Sonner?} to inclusion of lattice breaking translational invariance \cite{Horowitz-Tong,Donos,Ling-Tian}. }

Technically, most of those studies mainly rely on solving equations of motion (EOMs) for classical fields in AdS black hole, and particularly impose ingoing wave condition (for time-dependent problems) or regular condition (for static situations) near the black hole horizon. However, there is also possibility that a bulk field behaves as outgoing wave (Hawking radiation/mode) near the horizon. From the viewpoint of black hole physics, the ingoing wave condition captures dissipation, while the outgoing wave condition represents stochastic fluctuation or noise. To satisfy fluctuation-dissipation relations (FDRs) on the boundary, both ingoing mode and outgoing mode should be present for a bulk field \cite{Herzog:2002pc,deBoer:2008gu,Son:2009vu}. However, in AdS black hole with a single conformal boundary, inclusion of both ingoing and outgoing modes would inevitably result in infrared divergences \cite{deBoer:2008gu}. A self-consistent approach for curing this problem is recently proposed in \cite{Glorioso:2018mmw} (see \cite{Herzog:2002pc,Skenderis:2008dg,Skenderis:2008dh} for alternative approaches), which provides a holographic prescription for SK closed time contour for non-equilibrium system. Within this prescription, the holographic (radial) coordinate is complexified and analytically continued around the event horizon, forming a doubled Schwarzschild-AdS black hole with two conformal boundaries. In the past two years, this prescription was used to derive effective action for simple holographic systems \cite{Chakrabarty:2019aeu,Jana:2020vyx,Chakrabarty:2020ohe,Loganayagam:2020eue,Loganayagam:2020iol,Ghosh:2020lel,Bu:2020jfo}, which essentially involves solving {\it linear} EOMs in doubled Schwarzschild-AdS geometry.

In this work, we will adopt the holographic prescription of \cite{Glorioso:2018mmw} and construct the low energy EFT action of the holographic superconductor model \cite{Herzog:2010vz}. As an initial study, we will focus on the real-time dynamics of a fluctuating order parameter\footnote{In the probe limit, non-equilibrium nature of our study is reflected in this {\it real-time dynamics} of a fluctuating condensate.}: the dynamics of $U(1)$ charge diffusion gets decoupled via switching off spatial dependence. Compared to \cite{Chakrabarty:2019aeu,Jana:2020vyx,Chakrabarty:2020ohe,Loganayagam:2020eue,Loganayagam:2020iol,
Ghosh:2020lel,Bu:2020jfo}, the present study goes beyond linear approximation and tests validity of the prescription \cite{Glorioso:2018mmw} for nonlinear problems. The main goal is to introduce time-dependence into $\mathcal F_{\rm GL}$ \eqref{F_GL} and put it into a non-equilibrium QFT framework. In this way, both dissipation and fluctuation are {\it systematically} included. Near the critical point, we are able to (semi-)analytically derive the boundary effective action up to quartic order in the order parameter and first order in time-derivative. Our results share certain common features with the weakly coupled ones \cite{Kamenev2011,Levchenko2007}, which reflects the fact that both models are of mean field type and belong to the same model according to Hohenberg-Halperin classification for dynamical critical phenomena \cite{RevModPhys.49.435}. However, our results contain more complete structures in the quartic terms that are not fully explored in the weakly coupled ones \cite{Kamenev2011,Levchenko2007}.

The rest of this paper will be structured as follows. In section \ref{holo_setup} we present the holographic setup. In section \ref{solve_bulk_eom} we solve the bulk dynamics on the holographic SK contour. In section \ref{effective_action} we present the results for the time-dependent Ginzburg-Landau action in the spatially homogeneous limit. In section \ref{summary} we make a summary and outlook some future directions. The appendex \ref{boundary_action_derivation} supplements formal derivation of boundary effective action based on partially on-shell procedure in the bulk. The appendix \ref{sources} summarizes the source terms for perturbative bulk EOMs.
%
%The appendix \ref{cancellation_quadratic} provides further calculational details regarding the gauge field's kinetic term.

\iffalse

Holography has been quite useful in understanding dynamics of phase transition, e.g., chiral transition relevant in deconfined phase of QCD, superconducting phase transition in CMP, etc. In the beginning, these studies tried to understand the static properties in the symmetry-broken phase or condensed phase. critical exponents calculations (Zong-Gao-Zeng, some japanese authors, ...). Here, Ginzburg-Landau free energy was reconstructed ??? (Hou-Yin),

There are also some interesting works on exploring dynamic process of the phase transition, see 1005.0633, 1406.2329, and 1407.1862

Some latest progress on holographic superfluid model, vortex dynamics, time-dependent configurations, Kibble-Zurek mechanism etc.

While in the linear approximation, the fluctuation effects can be captured by FDT, but beyond linear regime, more interesting/important effects would appear.

For finite temperature systems, these involve both dissipation and fluctuations. More importantly, thermal fluctuation effects are crucial in correctly understanding physics near the phase transition critical point, see numeric simulations in which random fluctuations are put by hand in 1406.2329, and 1407.1862.

\fi

\section{The holographic model} \label{holo_setup}

In the probe limit, i.e. without considering the backreaction of matter fields, a holographic model for $s$-wave superconductor is simply the scalar QED in Schwarzschild-AdS geometry \cite{Gubser:2008px,Hartnoll:2008vx}. We will consider a five dimensional holographic superconductor model \cite{Horowitz:2008bn}
\begin{align}
S_0=\int d^5 x \sqrt{-g} \left[-\frac{1}{4} F_{MN} F^{MN} - (D_M \Psi)^*(D^M\Psi) - m_0^2 \Psi^* \Psi\right]. \label{S0}
\end{align}
Here, $F_{MN}= \nabla_M A_N - \nabla_N A_M$ and $D_M= \nabla_M- iq A_M$. Since a $U(1)$ gauge symmetry in the bulk corresponds to a global $U(1)$ symmetry on the boundary, the model \eqref{S0} indeed realizes superfluidity. The spontaneously breaking of $U(1)$ symmetry on the boundary is realized as formation of a charged scalar hair of the AdS black hole. In the ingoing Eddington-Finkelstein (EF) coordinate system $x^M= (r,v,x^i)$, the metric of Schwarzschild-AdS is given by the line element:
\begin{align}
ds^2= g_{MN} dx^M dx^N = 2dvdr- r^2f(r) dv^2 +r^2 \delta_{ij}dx^i dx^j, \qquad i,j=1,2,3, \label{EF_metric}
\end{align}
where $f(r)=1-r^4_h/r^4$, and the horizon radius $r_h$ determines the black hole temperature $T= r_h/\pi$. In the Schwarzschild coordinate system $\tilde x^M= (r,t,x^i)$, \eqref{EF_metric} changes as
\begin{align}
ds^2= \tilde g_{MN} d \tilde x^M d \tilde x^N = \frac{dr^2}{r^2f(r)} - r^2f(r) dt^2 +r^2 \delta_{ij}dx^i dx^j, \qquad i,j=1,2,3, \label{Schw_metric}
\end{align}

\begin{figure}[htbp]
\centering
\includegraphics[width=0.8\textwidth]{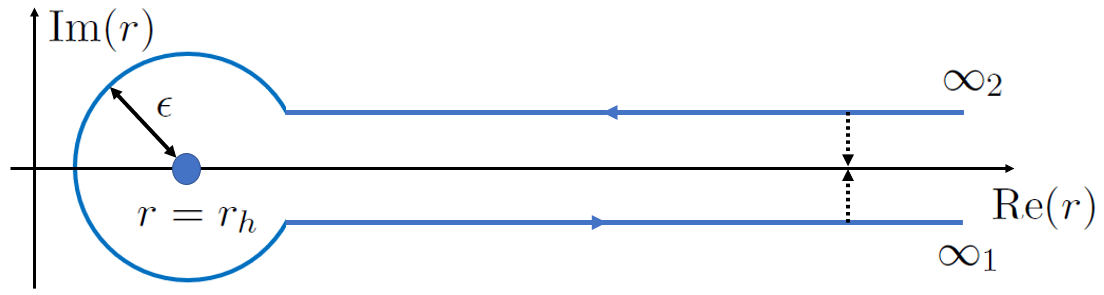}
\caption{The holographic prescription for the Schwinger-Keldysh closed time path \cite{Glorioso:2018mmw}: complexified radial coordinate and analytical continuation around the horizon $r_h$.}
\label{holographic_SK_contour}
\end{figure}

In order to incorporate both fluctuation and dissipation, the boundary theory should be put on the Schwinger-Keldysh (SK) time contour \cite{Kamenev2011}. A holographic dual for the SK time contour is proposed in \cite{Glorioso:2018mmw}, which complexifies the radial coordinate $r$ of \eqref{EF_metric} and analytically continues it around the event horizon $r=r_h$, see Figure \ref{holographic_SK_contour}. {Thanks to the probe approximation, the bulk metric is static so that stress tensor on the boundary does not depend on spacetime coordinate. While, going beyond probe limit, we will see non-equilibrium feature in the holographic contour (particularly, $r_h$ will be time-dependent), it is beyond the scope of present work and will be addressed elsewhere following the treatments of \cite{Crossley:2015tka,deBoer:2015ija}.}
%\sout{With the radial coordinate $r$ complexified, one should be careful about the meaning of complex conjugate, particularly, the kinetic term for the complex scalar field should be understood as
%\begin{align}
%(D_M\Psi)^* D^M\Psi = (\nabla_M \Psi^*+ iq A_M \Psi^*) (\nabla^M \Psi - iq A^M \Psi),
%\end{align}
%where $\Psi^* \equiv \Psi^*(r,x^\mu) \neq [\Psi(r,x^\mu)]^*$.}
    %{{I think we can defer discussion on complex conjugate for later. At this stage, the problem hasn't emerge yet.}}

From the variational problem of \eqref{S0}, we obtain bulk equations of motion (EOMs)
\begin{align}
&EA^N \equiv \nabla_M F^{MN}- iq \left[\Psi^* D^N \Psi- \Psi (D^N\Psi)^*\right]=0, \nonumber \\
&E\Psi \equiv D_M D^M \Psi- m_0^2 \Psi=0, \nonumber \\
&E\Psi^* \equiv (D_M D^M \Psi)^*- m_0^2 \Psi^*=0. \label{eom_EF}
\end{align}
The Maxwell equations could be further split into dynamical equations ($EA^\mu=0$) and constraint one ($EA^r=0$), where the latter $EA^r=0$ gives rise to current conservation equation for the boundary $U(1)$ current.

In the Schwarzschild coordinate system, the bulk EOMs are
\begin{align}
&E\tilde A^N \equiv \tilde \nabla_M \tilde F^{MN}- iq \left[\tilde \Psi^* \tilde D^N \tilde \Psi- \tilde \Psi (\tilde D^N \tilde \Psi)^*\right]=0, \nonumber \\
&E\tilde \Psi \equiv \tilde D_M \tilde D^M \tilde \Psi- m_0^2 \tilde \Psi=0, \nonumber \\
&E\tilde \Psi^* \equiv (\tilde D_M \tilde D^M \tilde \Psi)^*- m_0^2 \tilde \Psi^*=0, \label{eom_Schw}
\end{align}
where a tilde is to denote quantity in the Schwarzschild coordinate system. Going from \eqref{EF_metric} to \eqref{Schw_metric}, the two set of bulk EOMs \eqref{eom_EF} and \eqref{eom_Schw} are related as
\begin{align}
& EA^r =0 \Leftrightarrow  E\tilde A^r=0, \qquad  EA^v - \frac{EA^r}{r^2f(r)}=0 \Leftrightarrow E\tilde A^t=0, \qquad EA^k =0 \Leftrightarrow  E\tilde A^k=0 \nonumber \\
& E\Psi=0 \Leftrightarrow E\tilde\Psi=0, \qquad E\Psi^* =0 \Leftrightarrow E\tilde \Psi^*=0.
\end{align}
We will take a Schwarzschild radial gauge choice:
\begin{align}
\tilde A_r =0 \Longleftrightarrow A_r =- \frac{A_v}{r^2f(r)}. \label{Schw_radial_gauge}
\end{align}
In this gauge, the dynamical EOMs for components of $A_\mu$ and scalar fields are given by
\begin{align}
EA^v - \frac{EA^r}{r^2f(r)}=0, \qquad   EA^k =0, \qquad E\Psi=0, \qquad E\Psi^* =0. \label{eom_dynamic_EF}
\end{align}
Alternatively, if we had taken the EF radial gauge $A_r=0$, we should choose a different set of dynamical EOMs for $A_\mu$ instead:
\begin{align}
EA^v =0, \qquad   EA^k =0, \qquad E\Psi=0, \qquad E\Psi^* =0. \label{eom_dynamic_EF_alternative}
\end{align}

Near the AdS boundaries, the bulk action $S_0$ \eqref{S0} contains UV divergences, which can be removed by supplementing appropriate boundary terms. Indeed, those boundary terms should also guarantee that the bulk variational problem is well-posed. We will take the scalar mass as $m_0^2=-4$ saturating the Breitenlohner-Freedman bound. Near the AdS boundary $r=\infty_s ~ (s=1,2)$, the bulk fields behave as
\begin{align}
& A_\mu(r\to\infty_s) = \mathcal A_{s\mu}+ \frac{\partial_v \mathcal A_{s\mu}}{r} - \frac{1}{2}\partial^\nu \mathcal F_{s\mu\nu} \frac{\log r}{r^2} + \frac{\mathcal J_{s\mu}}{r^2} + \cdots, \nonumber \\
& A_r(r\to\infty_s) = - \frac{\mathcal A_{sv}}{r^2}+ \cdots, \nonumber \\
& \Psi(r\to\infty_s)= \psi_{bs} \frac{\log r}{r^2} + \frac{\Delta_s}{r^2}+ \cdots, \nonumber \\
& \Psi^*(r\to \infty_s)= \bar\psi_{bs} \frac{\log r}{r^2} + \frac{\bar \Delta_s}{r^2}+ \cdots,  \label{asymp_expand_contour}
\end{align}
where $\bar\psi_{bs}$ ($\bar \Delta_s$) is not necessarily the complex conjugate of $\psi_{bs}$ ($\Delta_s$), since $\Psi$ and $\Psi^*$ are two independent fields. $\mathcal F_{s\mu\nu}$ is the field strength of the external gauge potential $\mathcal A_{s\mu}$ for the boundary theory. Since $-d^2/4\le m^2 \le -d^2/4+1$ for $d=4$, the two modes of $\Psi$ ($\Psi^*$) are normalizable, and have conformal dimension two. Moreover, these two modes are related to each other via a canonical transformation \cite{Klebanov:1999tb}.
%\MF{The order parameter on the boundary and $\psi_{bs}(\bar{\psi}_{bs}$) have conformal dimension two. Since $-d^2/4\le m^2 \le -d^2/4+1$ for $d=4$, these modes of $\Psi(\Psi^*)$ are normalizable.  Two modes of a complex field are related each other via a canonical transformation~\cite{Klebanov:1999tb}. }
Thus, there are two quantization schemes for the scalar operator dual to the bulk scalar field. If $\psi_{bs}, \bar\psi_{bs}$ are taken as sources, the correct boundary terms would be
\begin{align}
& S_{\rm bdy}^{\rm I} = \int d^4x \sqrt{-\gamma} \mathcal{L}_{\rm bdy}^{\rm I} \bigg|_{r=\infty_1} - \int d^4x \sqrt{-\gamma} \mathcal{L}_{\rm bdy}^{\rm I} \bigg|_{r=\infty_2},  \nonumber \\
& \mathcal{L}_{\rm bdy}^{\rm I} = \frac{1}{4} F_{\mu\nu}F^{\mu\nu} \log r -2\Psi^* \Psi + \frac{\Psi^* \Psi}{\log r}, \label{Sbdy1}
\end{align}
where $\gamma$ is determinant of the induced metric on the AdS boundary. On the other hand, if $\Delta, \bar \Delta$ are taken as sources, the correct boundary terms would be
\begin{align}
& S_{\rm bdy}^{\rm II} = \int d^4x \sqrt{-\gamma} \mathcal{L}_{\rm bdy}^{\rm II} \bigg|_{r=\infty_1} - \int d^4x \sqrt{-\gamma} \mathcal{L}_{\rm bdy}^{\rm II} \bigg|_{r=\infty_2}, \nonumber \\
& \mathcal{L}_{\rm bdy}^{\rm II}= \frac{1}{4} F_{\mu\nu} F^{\mu\nu} \log r + 2\Psi^* \Psi - \frac{\Psi^* \Psi}{\log r} + n_M (\Psi^*\nabla^M \Psi + \Psi \nabla^M \Psi^* ), \label{Sbdy2}
\end{align}
where $n_M$ is the out-pointing unit normal vector of the AdS boundary. The last term of $\mathcal{L}^{\rm II}_{\rm bdy}$ is analogous to the counter terms given in \cite{Hartnoll:2008kx}.
Based on \eqref{asymp_expand_contour}, it is straightforward to check that, exactly on the AdS boundary, the bulk variational problem is well-defined:
\begin{align}
& \delta(S_0+S_{\rm bdy}^{\rm I}) = \int d^4x \left[\left( \bar \Delta_1 \delta \psi_{b1} + \Delta_1 \delta \bar \psi_{b1} + \check {\mathcal J}^\mu_1 \delta \mathcal A_{1\mu} \right) - \left( \bar \Delta_2 \delta \psi_{b2} + \Delta_2 \delta \bar \psi_{b2} + \check {\mathcal J}^\mu_2 \delta \mathcal A_{2\mu} \right) \right], \nonumber \\
& \delta(S_0+S_{\rm bdy}^{\rm II}) = \int d^4x \left[ \left( \bar \psi_{b1} \delta \Delta_1 + \psi_{b1} \delta \bar \Delta_1 + \check {\mathcal J}^\mu_1 \delta \mathcal A_{1\mu} \right) - \left( \bar \psi_{b2} \delta \Delta_2 + \psi_{b2} \delta \bar \Delta_2 + \check {\mathcal J}^\mu_2 \delta \mathcal A_{2\mu} \right) \right],
\end{align}
where the boundary current $\check {\mathcal J}^\mu$ is identical to the normalizable mode $\mathcal J^\mu$ up to contact terms.
Indeed, for a real scalar (with the same mass) in AdS$_5$, holographic renormalization was initially considered in \cite{Bianchi:2001kw} (see (5.30) therein), concluding the same scalar part of \eqref{Sbdy1}. In contrast, the boundary term proposed by Herzog \cite{Herzog:2010vz} is different from both \eqref{Sbdy1} and \eqref{Sbdy2}. In practical calculations, we will take the second quantization scheme \eqref{Sbdy2} so that boundary effective action is a functional of $\Delta, \bar \Delta$, which can be fluctuating. This is in the same spirit of the treatment of \cite{Ghosh:2020lel}.

Derivation of hydrodynamic effective action from AdS gravity has been nicely formulated in \cite{Crossley:2015tka} (see also \cite{Glorioso:2018mmw,Bu:2020jfo} for charge diffusion problem), based on early attempts \cite{Heemskerk:2010hk,Faulkner:2010jy,Nickel:2010pr}. The basis is the Gubser-Klebanov-Polyakov-Witten (GKPW) prescription \cite{Gubser:1998bc,Witten:1998qj} for AdS/CFT correspondence, which equals the partition functions of AdS gravity and dual CFT:
\begin{align}
Z_{\rm CFT} = Z_{\rm AdS}.
\end{align}
The CFT partition function $Z_{\rm CFT}$ could be written as a path integral over slow modes in the low energy EFT (collectively denoted by $X$):
\begin{align}
Z_{\rm CFT}= \int [DX] e^{i S_{eff}[X]}. \label{Z_CFT}
\end{align}
The AdS partition function $Z_{\rm AdS}$ is a path integral over bulk fields,
\begin{align}
Z_{\rm AdS}= \int DA_M^\prime D \Psi^\prime D\Psi^{*\prime} e^{i S_0 + i S_{\rm bdy}^{\rm II} }, \label{Z_AdS}
\end{align}
which will be computed in the saddle point approximation. In \eqref{Z_AdS} the primed configuration $(A_M^\prime, \Psi^\prime, \Psi^{*\prime})$ does not assume any gauge-fixing.

Apparently, the task of obtaining $S_{eff}$ from bulk theory amounts to identifying slow modes of low energy EFT, whose holographic dual should be kept off-shell on the gravity side \eqref{Z_AdS}. This has been nicely elaborated by Nickel and Son \cite{Nickel:2010pr} via re-examining the bulk $U(1)$ gauge symmetry. Instead of directly taking a specific gauge convention (e.g., radial gauge choice), the authors of \cite{Nickel:2010pr} achieved this by gauge transformation over a given configuration of bulk fields $(A_M^\prime, \Psi^\prime, \Psi^{*\prime})$. Consequently, the low energy dynamical variable associated with boundary $U(1)$ charge is identified with boundary value of the gauge transformation parameter $\Lambda$ \cite{Nickel:2010pr}.
%
%A specific gauge convention (e.g., radial gauge choice) could be achieved through a proper gauge transformation over :
%\begin{align}
%A_M^\prime \to A_M = A_M^\prime + \partial_M \Lambda, \qquad \Psi^\prime \to \Psi = \Psi^\prime e^{iq \Lambda}, \qquad  \Psi^{*\prime} \to \Psi^{*\prime} e^{-iq \Lambda}, \label{gauge_trans1}
%\end{align}
% On the AdS boundary, the gauge transformation \eqref{gauge_trans1} reads
%\begin{align}
%\mathcal A_\mu^\prime \to \mathcal A_\mu = \mathcal A_\mu^\prime + \partial_\mu \varphi, \qquad \Delta^\prime \to \Delta= \Delta^\prime e^{iq \varphi}, \qquad \bar\Delta^\prime \to \bar\Delta = \bar\Delta^\prime e^{-iq \varphi},
%\end{align}
%where $\varphi(x^\alpha) \equiv \Lambda(r=\infty,x^\alpha)$ is identified with the dynamical variable for the charge diffusion \cite{Nickel:2010pr}.
%
Therefore, going from field configuration $(A_M^\prime, \Psi^\prime, \Psi^{*\prime})$ without any gauge-fixing to the gauge-fixed one, say $(A_r=0, A_\mu, \Psi, \Psi^*)$ or $(A_r =- A_v/(r^2f(r)), A_\mu, \Psi, \Psi^*)$, is equivalent to changing integration variable in \eqref{Z_AdS} from $A_r^\prime$ to $\Lambda$. Meanwhile, in the saddle point approximation, in order to guarantee $\varphi$ to be off-shell, we shall not impose radial component (constraint equation) of Maxwell equation. Thus, with the dynamical EOMs solved only, the AdS partition function is eventually cast into
\begin{align}
Z_{\rm AdS} = \int D \varphi D \Delta D \bar\Delta e^{iS_0|_{\rm p.o.s} + iS_{\rm bdy}^{\rm II}},
\end{align}
which is the desired form for boundary EFT. Here, $S_0|_{\rm p.o.s}$ stands for the partially on-shell bulk action, obtained by plugging the solution for dynamical EOMs into $S_0$. Eventually, the boundary effective action is identified with the renormalized partially on-shell bulk action
\begin{align}
S_{eff}= S_0|_{\rm p.o.s} + S_{\rm bdy}^{\rm II}. \label{S_eff_S0_pos}
\end{align}
In practical calculation, $S_0|_{\rm p.o.s}$ will be obtained in a specific gauge, which does not generate ambiguity or violation of bulk gauge invariance. In appendix \ref{boundary_action_derivation}, starting from bulk path integral \eqref{Z_AdS}, we demonstrate as long as dynamical EOMs are correctly taken (in compatible with a specific gauge choice), the partially on-shell bulk action $S_0|_{\rm p.o.s}$ computed in different gauge choices takes the same form in terms of slow modes.

Such an off-shell procedure allows possible violation of current conservation by fluctuations, which is an essential ingredient of effective action. Note that, with a specific gauge choice, the dynamical EOMs can fully determine profiles of the bulk fields, given sufficient boundary conditions. This approach was first used to resum all-order derivatives in fluid-gravity correspondence \cite{Bu:2014sia,Bu:2014ena,Bu:2015ika}, and also employed to derive hydrodynamic effective action from gravity \cite{Crossley:2015tka,deBoer:2015ija,Glorioso:2018mmw,deBoer:2018qqm}. Nowadays, this approach is referred to as off-shell holography, or more precisely, partially on-shell holography.

For later convenience, we simplify the bulk action $S_0$ by using the EOMs for scalar fields. Upon integration by parts over the scalar's kinetic terms, the bulk action $S_0$ \eqref{S0} becomes
\begin{align}
S_0
%=& \int d^4x \int_{\infty_2}^{\infty_1} dr \sqrt{-g} \left[- \frac{1}{4} F_{MN} F^{MN} - \frac{1}{2}\nabla_M (\Psi^* D^M \Psi) - \frac{1}{2}\nabla_M [\Psi (D^M \Psi)^*] \right] \nonumber \\
%&+ \int d^4x \int_{\infty_2}^{\infty_1} dr \left\{ \frac{1}{2}\Psi[(D_M D^M \Psi)^*- m_0^2 \Psi^*]  + \frac{1}{2}\Psi^*[(D_M D^M \Psi)- m_0^2 \Psi]\right\}, \nonumber \\
=& -\int d^4x \sqrt{-\gamma} n_M \left( \frac{1}{2}\Psi^* D^M \Psi + \frac{1}{2}\Psi (D^M \Psi)^* \right) \bigg|_{r=\infty_2}^{r=\infty_1} \nonumber \\
& - \int d^4x \int_{\infty_2}^{\infty_1} dr \sqrt{-g} \frac{1}{4}F_{MN} F^{MN}, \label{S0_pos_PsiEOM}
\end{align}
where we made use of scalar's EOMs.

In \eqref{S0_pos_PsiEOM}, we do not make use of dynamical equations for $A_\mu$. With the help of dynamical EOMs \eqref{eom_dynamic_EF} and the gauge choice \eqref{Schw_radial_gauge}, the partially on-shell bulk action $S_0$ turns into
\begin{align}
S_0= & -\int d^4x \sqrt{-\gamma} n_M \left(\frac{1}{2} A_N F^{MN} + \frac{1}{2}\Psi^* D^M \Psi + \frac{1}{2}\Psi (D^M \Psi)^* \right) \bigg|_{r=\infty_2}^{r=\infty_1} \nonumber \\
& + \int d^4x \int_{\infty_2}^{\infty_1} dr \sqrt{-g} \frac{1}{2}A_N \nabla_M F^{MN} \nonumber \\
=& -\int d^4x \sqrt{-\gamma} n_M \left(\frac{1}{2} A_N F^{MN} + \frac{1}{2} \Psi^* D^M \Psi + \frac{1}{2} \Psi (D^M \Psi)^* \right) \bigg|_{r=\infty_2}^{r=\infty_1} \nonumber \\
& + \int d^4x \int_{\infty_2}^{\infty_1} dr \sqrt{-g} \frac{1}{2}iq A_N [\Psi^* (D^N \Psi) - \Psi (D^N \Psi)^*].  \label{S0_pos_Schw_radial}
\end{align}

While \eqref{S0_pos_PsiEOM} and \eqref{S0_pos_Schw_radial} are equivalent once partially on-shell bulk solutions are plugged in, we find \eqref{S0_pos_PsiEOM} is more convenient for practical calculations.

% I still don't understand why it is necessary. I guess the tricky part comes from using EOMs for gauge fields. How about not using EOMs for gauge fields? This has two benefits: making the evaluation of Maxwell term easier; the gauge invariance is manifestly valid}

\section{Perturbative solutions to dynamical EOMs} \label{solve_bulk_eom}

As announced before, we consider a spatially-homogeneous case
\begin{align}
A=- \frac{A_v(r,v)}{r^2f(r)}dr + A_v(r,v)dv, \qquad \Psi=\Psi(r,v), \qquad \Psi^*= \Psi^*(r,v), \label{ansatz}
\end{align}
which is a consistent ansatz. Physically, this would mean that the charge diffusion will be decoupled, and the focus is on fluctuation effects of the homogeneous scalar condensate and charge density.

Explicitly, the dynamical EOMs \eqref{eom_dynamic_EF} read
\begin{align}
0= & \partial_r (r^3 \partial_r A_v) + \left[ \frac{2r}{f(r)} \partial_r + \frac{1}{f(r)} - \frac{r f^{\prime}(r)}{f^2(r)} \right] \partial_v A_v
+ \frac{\partial_v^2 A_v}{rf^2(r)} \nonumber \\
& - \frac{iq r}{f(r)} \left(\Psi^* \partial_v \Psi - \Psi \partial_v \Psi^* \right)
-\frac{2q^2 r} {f(r)} \Psi^* \Psi A_v, \nonumber \\
0= & \partial_r[r^5f(r) \partial_r \Psi] + 2r^3 \partial_r \partial_v \Psi + 3r^2 \partial_v \Psi + \frac{2iqr}{f(r)} A_v \partial_v \Psi + \frac{iqr}{f(r)}\Psi \partial_v A_v \nonumber \\
& + \frac{q^2 r}{f(r)} A_v^2 \Psi -m_0^2 r^3 \Psi, \nonumber \\
0= & \partial_r[r^5f(r) \partial_r \Psi^*] + 2r^3 \partial_r \partial_v \Psi^* + 3r^2 \partial_v \Psi^* - \frac{2iqr}{f(r)} A_v \partial_v \Psi^* - \frac{iqr}{f(r)} \Psi^* \partial_v A_v  \nonumber \\
& + \frac{q^2 r}{f(r)} A_v^2 \Psi^* -m_0^2 r^3 \Psi^*. \label{EOM_homogeneous}
\end{align}
The constraint equation $EA^r =0$ is
\begin{align}
& 0= \partial_v\left[ \partial_r A_v + \frac{\partial_v A_v}{r^2f(r)} \right] - iq \left[ r^2f(r) \Psi^* \partial_r \Psi- r^2f(r) \Psi \partial_r \Psi^* + \Psi^* \partial_v \Psi - \Psi \partial_v \Psi^* \right]. \label{constraint_homogeneous}
\end{align}
Note that making the replacement $\Psi \to \Psi^*, q \to -q$ in the EOM for $\Psi$ gives rise to the EOM of $\Psi^*$.

\iffalse

{\color{red}

One would choose an alternative set of dynamical ODEs \eqref{eom_dynamic_EF_alternative}. Then, the time-component of the Maxwell equation becomes
\begin{align}
0= \partial_r (r^3 \partial_r A_v) + \partial_r\left[\frac{r\partial_v A_v}{f(r)}\right] + iq r^3 \left(\Psi^* \partial_r \Psi - \Psi \partial_r \Psi^* \right) -\frac{2q^2 r} {f(r)} \Psi^* \Psi A_v,
\end{align}
with the others unchanged.
}
\fi

As outlined in section \ref{holo_setup} (see also appendix \ref{boundary_action_derivation}), for the purpose of deriving boundary effective action, we will solve the dynamical EOMs \eqref{EOM_homogeneous}, leaving aside the constraint \eqref{constraint_homogeneous}. Here, we should specify suitable boundary conditions. For the time-dependent Ginzburg-Landau effective action, we freeze the fluctuation of gauge potentials, but allow the condensate to fluctuate. For $A_v$, we require
\begin{align}
A_v(r= \infty_1)= A_v(r=\infty_2)=\mu, \label{Av_AdS_condition}
\end{align}
which means there is no noise in the chemical potential, i.e., the difference $\mathcal{A}_{1v}-\mathcal{A}_{2v}$ vanishes.
For simplicity, we will assume chemical potential $\mu$ to be constant. However, under the conditions \eqref{Av_AdS_condition}, $A_v$ cannot be uniquely fixed. As explained in \cite{Glorioso:2018mmw}, one can additionally impose vanishing condition at the horizon
\begin{align}
A_v(r=r_h)=0.  \label{Av_horizon_condition}
\end{align}
For $\Psi$ and $\Psi^*$, we fix the scalar condensate, that is, we will take $\Delta_s,\bar \Delta_s$ as given, while $\psi_{bs},\bar \psi_{bs}$ will be functionals of $\Delta_s, \bar \Delta_s$ once \eqref{EOM_homogeneous} are solved over the radial contour. By physical considerations, we assume
\begin{align}
\bar \Delta_1=\Delta_1^*, \qquad \qquad \bar \Delta_2= \Delta_2^*, \label{real_Delta}
\end{align}
but $\bar \psi_{b1}$ ($\bar\psi_{b2}$) is in general not complex conjugate of $\psi_{b1}$ ($\psi_{b2}$), which will be clear later.

%
%More precisely, near the AdS boundaries $r=\infty_s$ ($s=1,2$), $\Psi, \Psi^*$ are expanded as
%\begin{align}
%\Psi(r\to \infty_s)= \psi_{bs} \frac{\log r}{r^2} + \frac{\Delta_s}{r^2} + \cdots, \qquad \Psi^*(r\to \infty_s)= \bar \psi_{bs} \frac{\log r}{r^2} + \frac{\bar \Delta_s}{r^2} + \cdots,
%\end{align}

We turn to solve the coupled nonlinear partial differential equations (PDEs) \eqref{EOM_homogeneous}. In general, it is challenging to obtain analytical solutions for \eqref{EOM_homogeneous}. We will search for perturbative schemes to simplify the nonlinear problem. First, we consider the hydrodynamic limit in which the system evolves slowly in time. Thus, the bulk fields $A_v$, $\Psi$ and $\Psi^*$ are expanded in powers of $\xi \sim \partial_v$:
\begin{align}
& A_v = A_v^{(0)} + \xi A_v^{(1)} + \cdots, \quad \Psi = \Psi^{(0)} + \xi \Psi^{(1)} + \cdots, \quad \Psi^* = \Psi^{*(0)} + \xi \Psi^{*(1)} +  \cdots. \label{derivative_expansion}
\end{align}
For our purpose, it is sufficient to truncate the derivative expansion \eqref{derivative_expansion} at the first order $\mathcal{O}(\xi^1)$. Furthermore, close to the critical point, the fluctuation of the scalar condensate is small. Consequently, at each order in the time-derivative expansion \eqref{derivative_expansion}, the bulk fields are further expanded in amplitude of the scalar condensate (i.e., $\lambda$-expansion):
\begin{align}
& A_v^{(l)} = A_v^{(l)(0)} + \lambda^2 A_v^{(l)(2)} + \lambda^4 A_v^{(l)(4)}+ \cdots,\nonumber \\
& \Psi^{(l)} = \lambda \Psi^{(l)(1)} + \lambda^3 \Psi^{(l)(3)} + \cdots, \nonumber \\
& \Psi^{*(l)} = \lambda \Psi^{*(l)(1)} + \lambda^3 \Psi^{*(l)(3)} + \cdots, \label{lambda-expansion}
\end{align}
where $\lambda \sim \Delta_s$.
Technically, the derivative expansion \eqref{derivative_expansion} renders the original system of nonlinear PDEs \eqref{EOM_homogeneous} into a system of nonlinear ODEs, which are further reduced into a system of linear ODEs by the $\lambda$-expansion \eqref{lambda-expansion}. For single AdS black hole, Herzog found that \cite{Herzog:2010vz} when the chemical potential takes a special value $\mu_0= 2 r_h$, those linear ODEs could be solved analytically if one further imposes regularity condition at the horizon. This special value $\mu_0$ corresponds to the critical point \cite{Herzog:2010vz}\footnote{Similarly, the critical temperature $T_c$ is expressed in terms of the density $\mathcal{J}_v$. Substituting $T=1/\pi$ and $(\mathcal{J}_v)_c =2$ into the relation $T/T_c=[(\mathcal{J}_{v})_c/\mathcal{J}_v]^{1/3}$, one obtains $T_c= 0.253 \mathcal{J}_v^{1/3}$, which agrees with $T_c$ for $m^2=-4$ in \cite{Horowitz:2008bn}.}.
 This motivates us to make a third expansion around the critical chemical potential $\mu_0$:
\begin{align}
&A_v^{(l)(m)}= A_v^{(l)(m)(0)} + \alpha A_v^{(l)(m)(1)} + \cdots, \nonumber \\
&\Psi^{(l)(m)}= \Psi^{(l)(m)(0)} + \alpha \Psi^{(l)(m)(1)} + \cdots, \nonumber \\
&\Psi^{*(l)(m)}= \Psi^{*(l)(m)(0)} + \alpha \Psi^{*(l)(m)(1)} + \cdots,
\end{align}
where $\alpha \sim \delta\mu$ with $\delta \mu$ a chemical potential perturbation.

Indeed, at each specific order in the triple expansion, the linear ODEs satisfied by these fields differ by the source terms:
\begin{align}
\Box_A A_v^{(l)(m)(n)}= j_v^{(l)(m)(n)}, \qquad \Box \Psi^{(l)(m)(n)} = j_\Psi^{(l)(m)(n)}, \qquad \Box \Psi^{*(l)(m)(n)} = j_{\Psi^*}^{(l)(m)(n)}, \label{eom_Box}
\end{align}
where
\begin{align}
\Box_A \bullet = \partial_r (r^3 \partial_r \bullet), \qquad \Box \bullet = \partial_r [r^5f(r) \partial_r \bullet] + \frac{q^2 r}{f(r)} (\bar A_v)^2 \bullet - m_0^2r^3 \bullet.
\end{align}
Here, $\bar A_v = \mu_0 (1-r_h^2/r^2)$ corresponds to the gauge potential on the critical point. $j_{v,\Psi,\Psi^*}^{(l)(m)(n)}$ are computed from fields of lower orders. The AdS boundary conditions summarized in \eqref{Av_AdS_condition} through \eqref{real_Delta} will be fully imposed on the lowest order fields in the triple expansion, so that the higher order fields will satisfy Dirichlet-like conditions at the AdS boundaries. All the fields belonging to the expansion of $A_v$ satisfy vanishing condition at the horizon.

The solution for $A_v$ can be obtained by direct integration over $r$. At the lowest order in $\xi$- and $\lambda$-expansions,
\begin{align}
A_v^{(0)(0)(0)} \equiv \bar A_v = \mu_0 \left(1-\frac{r_h^2}{r^2}\right), \qquad  A_v^{(0)(0)(1)} \equiv \delta \bar A_v= \delta\mu \left(1-\frac{r_h^2}{r^2}\right).
\end{align}
For higher order fields $(l+m>0)$, the solution for $A_v^{(l)(m)(n)}$ can be written in piecewise form,
\begin{align}
A_v^{(l)(m)(n)}= \int_{\infty_s}^r dr' \left[ \frac{1}{r^{\prime 3}} \int_{\infty_s}^{r^\prime} j_v^{(l)(m)(n)}(r^{\prime\prime}) dr'' + \frac{c_s^{(l)(m)(n)}}{r^{\prime3}} \right], \quad r\in [r_h, \infty_s), \quad s=1,2,
\end{align}
where the lower bound $\infty_s$ helps to distinguish between the upper branch $(s=2)$ and the lower branch $(s=1)$. The integration constant $c_s^{(l)(m)(n)}$ is determined by \eqref{Av_horizon_condition}:
\begin{align}
c_s^{(l)(m)(n)} = 2r_h^2 \int_{\infty_s}^{r_h} \frac{dr'}{r^{\prime 3}} \int_{\infty_s}^{r^\prime} j_v^{(l)(m)(n)}(r^{\prime\prime}) dr'' , \quad s=1,2. \label{c12lmn}
\end{align}
The normalizable modes $\mathcal J_{sv}$, cf. \eqref{asymp_expand_contour}, are
\begin{align}
\mathcal J_{sv}^{(l)(m)(n)}= - \frac{1}{2} c_s^{(l)(m)(n)}, \qquad s=1,2.
\end{align}
$\mathcal J_{sv}$ have the interpretation as charge density induced by fluctuation of scalar condensate.

The lowest order parts of $\Psi$ and $\Psi^*$, say $\Psi^{(0)(1)(0)}$ and $\Psi^{*(0)(1)(0)}$, obey homogeneous ODEs:
\begin{align}
\Box \Psi^{(0)(1)(0)} = \Box \Psi^{*(0)(1)(0)}=0.
\end{align}
In order to have analytical solutions, we will set $q\mu_0 = 2r_h$ and $m^2=-4$ as realized in \cite{Herzog:2010vz}.
Since we will impose the AdS boundary conditions \eqref{asymp_expand_contour} over $\Psi^{(0)(1)(0)}$ and $\Psi^{*(0)(1)(0)}$, the lowest order solutions are
\begin{align}
&\Psi^{(0)(1)(0)}= \frac{\Delta_2}{r_h^2+ r^2} - \frac{\Delta_1 - \Delta_2}{2i\pi} \frac{\log r- \log(r^2 - r_h^2)}{r_h^2 +r^2}, \nonumber \\
&\Psi^{*(0)(1)(0)}= \frac{\bar\Delta_2}{r_h^2+ r^2} - \frac{\bar\Delta_1 - \bar \Delta_2} {2i\pi} \frac{\log r- \log(r^2 - r_h^2)}{r_h^2 +r^2}. \label{Psi_010}
\end{align}
Clearly, $\Psi^{(0)(1)(0)}$ and $\Psi^{*(0)(1)(0)}$ are related by $\Delta_s\to\bar{\Delta}_s$ but not complex conjugate. The reason is not difficult to see: the symbol $*$ corresponds to charge conjugation, under which $\Delta_s\to\bar{\Delta}_s$. However, we have also complexified the radial coordinate, which leads to the factor {$\frac{1}{2i\pi}$} in the second terms. This factor should not flip sign under charge conjugation.
From \eqref{Psi_010}, we read off the lowest order results for $\psi_b$ and $\bar\psi_b$:
\begin{align}
\psi_{b1}^{(0)(1)(0)}= \psi_{b2}^{(0)(1)(0)}= \frac{\Delta_1 - \Delta_2}{2i\pi}, \qquad \bar\psi_{b1}^{(0)(1)(0)}= \bar \psi_{b2}^{(0)(1)(0)}= \frac{\bar \Delta_1 - \bar \Delta_2}{2i\pi}, \label{psi_b_010}
\end{align}
which obviously confirms that $\bar \psi_b \neq \psi_b^*$ even though $\bar \Delta = \Delta^*$.

At higher orders in the triple expansion of $\Psi$ and $\Psi^*$, we will solve the inhomogeneous ODEs using the Green's function method. The Green's function is defined as
\begin{align}\label{GRE319}
\Box G(r,r^\prime) = \delta (r-r^\prime), \qquad r\in (\infty_2, \infty_1),
\end{align}
where the right hand side is the delta function in terms of $r$ and $r'$.
The Green's function becomes unique if we further impose homogeneous boundary conditions
\begin{align}
G(r\to \infty_2, r^\prime) \to \#_2 \frac{\log r}{r^2} + \frac{0}{r^2} + \cdots, \qquad G(r\to \infty_1, r^\prime) \to \#_1 \frac{\log r}{r^2} + \frac{0}{r^2} + \cdots, \label{G_AdS_behavior}
\end{align}
where $\#_{1,2}$ are not constrained. Then, the solutions for $\Psi, \Psi^*$ at each order $(l+m>1)$ are
\begin{align}
& \Psi^{(l)(m)(n)}(r) = \int_{\infty_2}^{\infty_1} dr^{\prime} G(r,r^\prime) j_{\Psi}^{(l)(m)(n)}(r^\prime), \nonumber \\
& \Psi^{*(l)(m)(n)}(r) = \int_{\infty_2}^{\infty_1} dr^{\prime} G(r,r^\prime) j_{\Psi^*}^{(l)(m)(n)}(r^\prime). \label{Psi_Green_sol}
\end{align}

For the purpose of constructing the Green's function $G(r,r^\prime)$, we look for linearly independent basis solutions $u_1(r)$ and $u_2(r)$ for the homogeneous ODE:
\begin{align}
\Box u=0.
\end{align}
Moreover, we define the basis solutions by requiring each of $u_{1,2}$ obeys one of the homogenous AdS boundary conditions:
\begin{align}
& u_2(r\to \infty_2) \to \frac{\log r}{r^2} + \frac{0}{r^2} + \cdots, \qquad \quad u_2(r \to \infty_1) \to \frac{k_1}{r^2} \log r + \frac{k_2}{r^2}+ \cdots, \nonumber \\
& u_1(r\to \infty_2) \to \frac{k_3}{r^2} \log r + \frac{k_4}{r^2} + \cdots, \qquad u_1(r \to \infty_1) \to \frac{\log r}{r^2} + \frac{0}{r^2} + \cdots. \label{u1u2_AdS_condition}
\end{align}
Here, $k_{1,2,3,4}$ are unconstrained. The linearly independent basis solutions are uniquely fixed as
\begin{align}
u_2(r) = - \frac{\log r - \log(r^2-r_h^2)}{r_h^2+r^2}, \qquad  u_1(r) = - \frac{\log r - \log(r^2-r_h^2)}{r_h^2+r^2} - \frac{2i \pi}{r_h^2+r^2}. \label{u1u2}
\end{align}
Note that $u_1$ and $u_2$ are not regular at the black hole horizon unlike that in Poincar{\' e} AdS coordinates \cite{Gubser:2002zh}.
%with $\arg(r-r_h)=0$ for $r$ on the upper branch and $\arg(r-r_h)=2\pi$ for $r$ on the lower branch.
In terms of basis solutions $u_{1,2}(r)$, the lowest order fields \eqref{Psi_010} are
\begin{align}
\Psi^{(0)(1)(0)}(r)= \frac{\Delta_1}{2i \pi} u_2(r) - \frac{\Delta_2}{2i\pi} u_1(r), \qquad \Psi^{*(0)(1)(0)}(r) = \frac{\bar \Delta_1}{2i \pi} u_2(r) - \frac{\bar \Delta_2}{2i\pi} u_1(r).  \label{Psi_010_new}
\end{align}
Since $r$ varies along the radial contour of Figure \ref{holographic_SK_contour}, the logarithmic function $\log(r^2-r_h^2)$ is multi-valued, particularly, going from the $\infty_2$-branch ($r\in [r_h+ \epsilon, \infty_2)$) to the $\infty_1$-branch ($r\in [r_h+ \epsilon, \infty_1)$) this function will receive an extra piece $2i\pi$, as implied in the behavior of $u_1(r\to \infty_1)$. Because the right hand side of  \eqref{GRE319} is zero for $r\neq r'$, the Green's function is
\begin{align}
G(r,r^\prime) = \frac{1}{r^{\prime5} f(r^\prime) W(r^\prime)} \left[ u_2(r) u_1(r^\prime) \Theta(r^\prime - r) + u_1(r) u_2(r^\prime) \Theta(r-r^\prime) \right],
\end{align}
where $W(r)$ is the Wronskian determinant of $u_1(r)$ and $u_2(r)$:
\begin{align}
W(r) \equiv u_2(r) \partial_r u_1(r) - u_1(r) \partial_r u_2(r) = \frac{2i\pi}{r^5-r_h^4r}.
\end{align}
The step function $\Theta(r-r^\prime)$ is defined on the radial contour of Figure \ref{holographic_SK_contour}, with $r>r^\prime$ ($r<r^\prime$) understood as counter clockwise path-ordered relations. This property, combined with \eqref{u1u2_AdS_condition}, guarantees the asymptotic behavior for the Green's function \eqref{G_AdS_behavior}. It is also straightforward to check that \eqref{Psi_Green_sol} do satisfy the EOMs \eqref{eom_Box} with correct boundary conditions. Indeed, near the AdS boundaries, the Green's function $G$ behaves as
\begin{align}
G(r\to \infty_2, r^\prime) \to \frac{u_1(r^\prime)}{2i\pi} \frac{\log r}{r^2} + \frac{0}{r^2}, \qquad  G(r\to \infty_1, r^\prime) \to \frac{u_2(r^\prime)}{2i\pi} \frac{\log r}{r^2} + \frac{0}{r^2},
\end{align}
which helps to extract $\psi_b$ and $\bar\psi_b$ (cf. \eqref{asymp_expand_contour}):
\begin{align}
\psi_{b2}^{(l)(m)(n)} = \frac{1}{2i\pi}\int_{\infty_2}^{\infty_1} dr u_1(r) j_\Psi^{(l)(m)(n)}(r), \qquad \psi_{b1}^{(l)(m)(n)} = \frac{1}{2i\pi}\int_{\infty_2}^{\infty_1} dr u_2(r) j_\Psi^{(l)(m)(n)}(r), \nonumber \\
\bar \psi_{b2}^{(l)(m)(n)} = \frac{1}{2i\pi}\int_{\infty_2}^{\infty_1} dr u_1(r) j_{\Psi^*}^{(l)(m)(n)}(r), \qquad \bar \psi_{b1}^{(l)(m)(n)} = \frac{1}{2i\pi}\int_{\infty_2}^{\infty_1} dr u_2(r) j_{\Psi^*}^{(l)(m)(n)}(r). \label{psi_b}
\end{align}

Given the ansatz \eqref{ansatz}, the partially on-shell bulk action \eqref{S0_pos_PsiEOM} is
%\begin{align}
%S_0= & \int d^4x \left\{ \frac{r}{4f(r)} \partial_v (A_v^2) + \frac{1}{2}r^3 A_v \partial_r A_v - \Psi^* \left[ r^5f(r)\partial_r + r^3 \partial_v \right]\Psi \right\}\bigg|_{r=\infty_2}^{r=\infty_1} \nonumber \\
%& + \int d^4x \int_{\infty_2}^{\infty_1} dr \left\{ - \frac{r}{2f(r)} A_v \partial_v\partial_r A_v - \frac{1}{2}r^3 A_r \partial_v^2 A_r - \frac{q^2r}{f(r)} A_v^2 \Psi^* \Psi \right. \nonumber \\
%& \left. \qquad \qquad \qquad \qquad + \frac{1}{2}iq r^3 A_v (\Psi^* \partial_r \Psi - \Psi \partial_r \Psi^*)  \right\} \label{S0_pos3}
%\end{align}
%Here, we can make further simplifications. The asymptotic expansion \eqref{asymp_expand} helps to drop the first and last terms in the first line of \eqref{S0_pos3}. Moreover, we will keep up to first order in time derivative, so that the second term in the second line of \eqref{S0_pos3} would also be dropped. Eventually, \eqref{S0_pos3} turns into
\begin{align}
S_0= & \int d^4x \left[ - \frac{1}{2} \Psi^* r^5f(r)\partial_r \Psi - \frac{1}{2} \Psi r^5f(r)\partial_r \Psi^* \right]\bigg|_{r=\infty_2}^{r=\infty_1} \nonumber \\
& - \int d^4x \int_{\infty_2}^{\infty_1} dr \sqrt{-g} \frac{1}{4}F_{MN}F^{MN}, \label{S0_pos_explicit}
\end{align}
where we have dropped some terms that vanish explicitly in the limit $r\to \infty_{1,2}$. Since the goal is to derive $S_{eff}$ up to quartic order in the amplitude of the order parameter, we will truncate the triple expansion appropriately. The relevant source terms are summarized in appendix \ref{sources}.

%
%We will perform a triple expansion for bulk fields:
%
%\noindent $\bullet$ time-derivative expansion: $\partial_v\sim \xi$, up to $\mathcal{O}(\xi^1)$
%
%\noindent $\bullet$ expansion in terms of amplitude of scalar field: $\lambda$. Originally, we would want to do calculations up to $\mathcal{O}(\lambda^4)$, but it turns out to be very involved. We may truncate the expansion at $\mathcal{O}(\lambda^2)$.
%
%\noindent $\bullet$ expansion in terms of chemical potential correction $\delta \mu$, so that the total chemical potential $\mu = \mu_0(=2) + \delta \mu$. We may introduce another expansion parameter $\delta \mu \sim \alpha$. We will keep to $\mathcal{O}(\alpha^1)$. Physically, this correction will bring us a little bit away from the critical point (at which the phase transition happens).

\section{Time-dependent Ginzburg-Landau effective action} \label{effective_action}

In this section, based on the solutions obtained in section \ref{solve_bulk_eom}, we compute the partially on-shell bulk action \eqref{S0_pos_explicit}, giving rise to the boundary effective action $S_{eff} =S_0 + S_{\rm bdy}^{\rm II}$. Using the asymptotic expansion \eqref{asymp_expand_contour}, it is direct to check that the bulk pieces of \eqref{S0_pos_explicit} are finite, while UV divergences from the surface terms of \eqref{S0_pos_explicit} can be removed by $S_{\rm bdy}^{\rm II}$ \eqref{Sbdy2}:
\begin{align}
S_{eff} = & \int d^4 x \left[ \frac{1}{2}\left( - \psi_{b1} \Delta_1^* + \psi_{b2} \Delta_2^* - \bar \psi_{b1} \Delta_1 + \bar \psi_{b2} \Delta_2 \right) \right] \nonumber \\
& -\int d^4x \int_{\infty_2}^{\infty_1} dr \sqrt{-g} \frac{1}{4} F_{MN}F^{MN}. \label{S_tot}
\end{align}
Recall that in order to have analytical basis solutions $u_{1,2}$ \eqref{u1u2}, we have set $q\mu_0 = 2r_h$ and $m^2=-4$.
The charge $q$ will be absorbed into $A_v$. We will also set $r_h=1$ for convenience. The bulk integral of \eqref{S_tot} will be computed by splitting the radial contour of Figure \ref{holographic_SK_contour} into three parts:
\begin{align}
\int_{\infty_2}^{\infty_1} dr = \int_{\infty_2}^{r_h+\epsilon} dr + \int_{\mathcal C} i\epsilon e^{i\theta} d\theta  + \int_{r_h+\epsilon}^{\infty_1} dr, \label{contour_split}
\end{align}
where $\mathcal C$ denotes the infinitesimal circle of Figure \ref{holographic_SK_contour}.

Given the specific ansatz \eqref{ansatz}, the calculation of quadratic order action gets simplified accidentally. Schematically, $A$ or $F$ could be written as
\begin{align}
A= \bar A + \delta A, \qquad \qquad F = \bar F + \delta F,
\end{align}
where $\bar A = A^{(0)(0)(0)}, \bar F= F^{(0)(0)(0)}$, and $\delta A, \delta F$ denote all possible higher order corrections in the triple expansion. The gauge field's kinetic term is expanded as
\begin{align}
S_A &= - \frac{1}{4} \int d^4x \int_{\infty_2}^{\infty_1} dr \sqrt{-g} F^2 \nonumber \\
&= \int d^4x \int_{\infty_2}^{\infty_1} dr \sqrt{-g} \left[ -\frac{1}{4} \bar F^2 - \frac{1}{2} \bar F^{MN} \delta F_{MN} \right] + \mathcal O((\delta A)^2), \nonumber \\
&= \int d^4x \int_{\infty_2}^{\infty_1} dr \, 2\mu_0 r_h^2 (\partial_r \delta A_v - \partial_v \delta A_r) + \mathcal O((\delta A)^2), \label{cancellation_proof}
\end{align}
which obviously vanishes, up to total time-derivative terms and nonlinear terms. Here, we made use of the fact that $\delta A_v$ vanishes at the AdS boundaries and the horizon\footnote{The boundary condition at the horizon is also needed because we split the radial contour as \eqref{contour_split} and the contribution from the infinitesimal circle vanishes.}. %In appendix \ref{cancellation_quadratic}, we will confirm this observation via an alternative calculation for the gauge field's kinetic term in \eqref{S_tot} at the level of quadratic order in the order parameter.
Thanks to this accidental cancellation, at the quadratic order in $\Delta_s$ only scalar's surface terms make non-vanishing contributions to the boundary effective action. Moreover, this trick is very helpful in simplifying computations of quartic order action.

We find it more convenient to use a piecewise form for $\Psi^{(0)(1)(0)}$ of \eqref{Psi_010}
\begin{align}\label{PSI45}
& \Psi^{(0)(1)(0)}(r) = \frac{\Delta_2}{1+r^2} + \frac{\Delta_2 - \Delta_1}{2i\pi} \frac{\log r -\log(r^2-1)}{1+r^2},\qquad r\in [r_h+\epsilon, \infty_2), \nonumber \\
& \Psi^{(0)(1)(0)}(\theta) = \left [\frac{\Delta_2}{1+r^2} + \frac{\Delta_2 - \Delta_1}{2i\pi} \frac{\log r -\log(r^2-1)}{1+r^2} \right] \bigg|_{r=r_h+\epsilon e^{i\theta}},\qquad \theta\in [0, 2\pi], \nonumber \\
& \Psi^{(0)(1)(0)}(r) = \frac{\Delta_1}{1+r^2} + \frac{\Delta_2 - \Delta_1}{2i\pi} \frac{\log r -\log(r^2-1)}{1+r^2},\qquad r\in [r_h+\epsilon, \infty_1),
\end{align}
and similarly for $\Psi^{*(0)(1)(0)}$ of \eqref{Psi_010}:
\begin{align}\label{PSI46}
& \Psi^{*(0)(1)(0)}(r) = \frac{\Delta_2^*}{1+r^2} + \frac{\Delta_2^* - \Delta_1^*}{2i\pi} \frac{\log r -\log(r^2-1)}{1+r^2},\qquad r\in [r_h+\epsilon, \infty_2), \nonumber \\
& \Psi^{*(0)(1)(0)}(\theta) = \left [\frac{\Delta_2^*}{1+r^2} + \frac{\Delta_2^* - \Delta_1^*}{2i\pi} \frac{\log r -\log(r^2-1)}{1+r^2} \right] \bigg|_{r=r_h+\epsilon e^{i\theta}},\qquad \theta\in [0, 2\pi], \nonumber \\
& \Psi^{*(0)(1)(0)}(r) = \frac{\Delta_1^*}{1+r^2} + \frac{\Delta_2^* - \Delta_1^*}{2i\pi} \frac{\log r -\log(r^2-1)}{1+r^2},\qquad r\in [r_h+\epsilon, \infty_1).
\end{align}
Fields in \eqref{PSI45} and \eqref{PSI46} have a logarithmic divergence at the black hole horizon. Due to this divergence, these fields can satisfy two different boundary conditions at $\infty_{1,2}$. Otherwise, these fields become regular and $\Delta_1=\Delta_2(\bar{\Delta}_1=\bar{\Delta}_2)$.

In accord with the triple expansion of the bulk fields, the boundary effective Lagrangian $\mathcal L_{eff}$, defined by $S_{eff}= \int d^4 x \mathcal{L}_{eff}$, is expanded as
\begin{align}
\mathcal L_{eff}= \mathcal L_{eff}^{(0)(2)(0)} + \mathcal L_{eff}^{(0)(2)(1)} + \mathcal L_{eff}^{(0)(4)(0)} + \mathcal L_{eff}^{(1)(2)(0)} + \cdots.
\end{align}

$\bullet$  $\mathcal L_{eff}^{(0)(2)(0)}$

From the formal analysis \eqref{cancellation_proof}, at this order only the scalar's surface term survive:
\begin{align}
\mathcal L_{eff}^{(0)(2)(0)} = \frac{1}{2} \left[ -\psi_{b1}^{(0)(1)(0)} \Delta_1^* + \psi_{b2}^{(0)(1)(0)} \Delta_2^* -\bar \psi_{b1}^{(0)(1)(0)} \Delta_1 + \bar\psi_{b2}^{(0)(1)(0)} \Delta_2 \right]
\end{align}
which is easy to compute with the help of \eqref{psi_b_010}:
\begin{align}
\mathcal L_{eff}^{(0)(2)(0)} = - \frac{\Delta_1 - \Delta_2}{2i\pi} (\Delta_1^* - \Delta_2^*).
\end{align}
Introducing difference and average combinations as the $(r,a)$-basis:
\begin{align}
\Delta_a = \Delta_1 - \Delta_2, \qquad \Delta_r = \frac{1}{2}(\Delta_1 + \Delta_2),
\end{align}
the boundary effective Lagrangian $\mathcal L_{eff}^{(0)(2)(0)}$ read as
\begin{align}
\mathcal L_{eff}^{(0)(2)(0)} = \frac{i}{2\pi} \Delta_a^* \Delta_a. \label{Leff_020}
\end{align}

$\bullet$  $\mathcal L_{eff}^{(0)(2)(1)}$

We move on to the $\delta \mu$-correction, i.e., the term $\mathcal L_{eff}^{(0)(2)(1)}$, which just like $\mathcal L_{eff}^{(0)(2)(0)}$ requires to compute the scalar's surface term only:
\begin{align}
\mathcal L_{eff}^{(0)(2)(1)}= \frac{1}{2}\left[ -\psi_{b1}^{(0)(1)(1)} \Delta_1^* + \psi_{b2}^{(0)(1)(1)} \Delta_2^* -\bar \psi_{b1}^{(0)(1)(1)} \Delta_1 + \bar \psi_{b2}^{(0)(1)(1)} \Delta_2 \right].
\end{align}
The scalar's surface term involves $\psi_{bs}^{(0)(1)(1)}$ and $\bar\psi_{bs}^{(0)(1)(1)}$ ($s=1,2$), cf. \eqref{psi_b}:
\begin{align}
& \psi_{b1}^{(0)(1)(1)} = \frac{1}{2i\pi} \int_{\infty_2}^{\infty_1} dr u_2(r) \frac{-2q^2r}{f(r)} \bar A_v(r) \delta\bar A_v(r) \Psi^{(0)(1)(0)}(r), \nonumber \\
& \psi_{b2}^{(0)(1)(1)} = \frac{1}{2i\pi} \int_{\infty_2}^{\infty_1} dr u_1(r) \frac{-2q^2r}{f(r)} \bar A_v(r) \delta\bar A_v(r) \Psi^{(0)(1)(0)}(r), \nonumber \\
& \bar\psi_{b1}^{(0)(1)(1)} = \frac{1}{2i\pi} \int_{\infty_2}^{\infty_1} dr u_2(r) \frac{-2q^2r}{f(r)} \bar A_v(r) \delta\bar A_v(r) \Psi^{*(0)(1)(0)}(r), \nonumber \\
& \bar\psi_{b2}^{(0)(1)(1)} = \frac{1}{2i\pi} \int_{\infty_2}^{\infty_1} dr u_1(r) \frac{-2q^2r}{f(r)} \bar A_v(r) \delta\bar A_v(r) \Psi^{*(0)(1)(0)}(r),
\label{psi_011}
\end{align}
which are obtained by substituting the source terms \eqref{jPsi_011} into \eqref{psi_b}. By splitting the contour as in \eqref{contour_split}, \eqref{psi_011} can be evaluated to give
%are evaluated in the same way as \eqref{APsiPsi_bulk_020}. The results are
\begin{align}
& \psi_{b1}^{(0)(1)(1)}= \frac{\delta\mu}{\mu_0} \left[\frac{\log2}{i\pi}(\Delta_2- \Delta_1) -\Delta_1 \right], \qquad
\psi_{b2}^{(0)(1)(1)}= \frac{\delta\mu}{\mu_0} \left[\frac{\log2}{i\pi}(\Delta_2- \Delta_1) -\Delta_2 \right], \nonumber \\
& \bar\psi_{b1}^{(0)(1)(1)}= \frac{\delta\mu}{\mu_0} \left[\frac{\log2}{i\pi}(\Delta_2^* - \Delta_1^*) -\Delta_1^* \right], \qquad
\bar\psi_{b2}^{(0)(1)(1)}= \frac{\delta\mu}{\mu_0} \left[\frac{\log2}{i\pi}(\Delta_2^* - \Delta_1^* ) -\Delta_2^* \right].
\end{align}
At this order, the effective Lagrangian is
\begin{align}
\mathcal L_{eff}^{(0)(2)(1)} & = \delta\mu \left[ \frac{\log2}{i\pi}(\Delta_2^*- \Delta_1^*)(\Delta_2 - \Delta_1) - (\Delta_2^* \Delta_2 - \Delta_1^* \Delta_1) \right] \nonumber \\
& = \delta\mu \left[ \frac{\log2}{i\pi} \Delta_a^* \Delta_a +( \Delta_a \Delta_r^* + \Delta_a^* \Delta_r) \right]. \label{Leff_021}
\end{align}

$\bullet$  $\mathcal L_{eff}^{(1)(2)(0)}$

Here, we consider first order time-derivative correction $\mathcal L_{eff}^{(1)(2)(0)}$. Just like $\mathcal L_{eff}^{(0)(2)(0)}$ and $\mathcal L_{eff}^{(0)(2)(1)}$, only scalar's surface term contributes at this order:
\begin{align}
\mathcal L_{eff}^{(1)(2)(0)}= \frac{1}{2} \left[ -\psi_{b1}^{(1)(1)(0)} \Delta_1^* + \psi_{b2}^{(1)(1)(0)} \Delta_2^* - \bar \psi_{b1}^{(1)(1)(0)} \Delta_1 + \bar \psi_{b2}^{(1)(1)(0)} \Delta_2 \right]
\end{align}
The scalar's surface contribution involves:
\begin{align}
& \psi_{b1}^{(1)(1)(0)} = \frac{1}{2i\pi} \int_{\infty_2}^{\infty_1} dr u_2(r) j_{\Psi}^{(1)(1)(0)}(r), \nonumber \\
& \psi_{b2}^{(1)(1)(0)} = \frac{1}{2i\pi} \int_{\infty_2}^{\infty_1} dr u_1(r) j_{\Psi}^{(1)(1)(0)}(r), \nonumber \\
& \bar\psi_{b1}^{(1)(1)(0)} = \frac{1}{2i\pi} \int_{\infty_2}^{\infty_1} dr u_2(r) j_{\Psi^*}^{(1)(1)(0)}(r), \nonumber \\
& \bar\psi_{b2}^{(1)(1)(0)} = \frac{1}{2i\pi} \int_{\infty_2}^{\infty_1} dr u_1(r) j_{\Psi^*}^{(1)(1)(0)}(r),
\end{align}
where the source terms $j_{\Psi}^{(1)(1)(0)}$ and $j_{\Psi^*}^{(1)(1)(0)}$ presented in \eqref{jPsi_110} are known analytically. While the computation of this term will be analogous to \eqref{psi_011}, the integral along the infinitesimal circle does not vanish when $\epsilon \to 0$. For illustration, we take the computation of $\psi_{b1}^{(1)(1)(0)}$ as an example. In order to calculate the integral along the infinitesimal circle, we need near-horizon behavior of the integrand:
\begin{align}
u_2(r) j_\Psi^{(1)(1)(0)}(r) \xrightarrow[]{r\to 1} \frac{i\partial_v(\Delta_1 - \Delta_2)} {4\pi (r-1)} \left[ \log2 + \log(r-1) \right] + \cdots,
\end{align}
where the $\cdots$ will not make finite contribution to $S_{eff}$ once $\epsilon \to 0$.
So, the integral along the infinitesimal circle is
\begin{align}
\int_{\mathcal C} dr u_2(r) j_\Psi^{(1)(1)(0)}(r) = \frac{1}{2}\partial_v(\Delta_2 - \Delta_1) \log\epsilon + \frac{\partial_v(\Delta_2 - \Delta_1)}{2} (\log2 + i\pi) + \cdots.
\end{align}
Next, we consider the contribution from the upper and lower horizontal legs:
\begin{align}
& \int_{1+\epsilon}^{\infty_1} dr u_2(r) j_\Psi^{(1)(1)(0)}(r) - \int_{1+\epsilon}^{\infty_2} dr u_2(r) j_\Psi^{(1)(1)(0)}(r) \nonumber \\
= & \int_{1+\epsilon}^{\infty} dr \left\{ \left[u_2(r) j_\Psi^{(1)(1)(0)}(r) \right] \bigg|_{r\in[1+\epsilon, \infty_1)} - \left[u_2(r) j_\Psi^{(1)(1)(0)}(r) \right] \bigg|_{r\in[1+\epsilon, \infty_2)} \right\} \nonumber \\
= & \frac{1}{2} \partial_v(\Delta_1 - \Delta_2) \log\epsilon + \frac{1}{2}(1-i) \left[(1+2i) \pi \partial_v \Delta_1 - \log2 \, \partial_v (\Delta_2 - \Delta_1) \right]+ \cdots,
\end{align}
where the logarithmic divergence is exactly cancelled by the logarithmic divergence from the infinitesimal circle. The rest $\psi_b$'s could be calculated in the same fashion. The results are
\begin{align}
&\psi_{b1}^{(1)(1)(0)} = \frac{\log2}{4\pi}\partial_v(\Delta_2- \Delta_1) +\left(- \frac{3i}{4}\partial_v \Delta_1 + \frac{1}{4} \partial_v \Delta_2\right), \nonumber \\
&\psi_{b2}^{(1)(1)(0)}= \frac{\log2}{4\pi}\partial_v(\Delta_2- \Delta_1) +\left(- \frac{3i}{4}\partial_v \Delta_2 + \frac{1}{4} \partial_v \Delta_1\right), \nonumber \\
&\bar\psi_{b1}^{(1)(1)(0)} = \frac{\log2}{4\pi} \partial_v(\Delta_1^* - \Delta_2^*) + \left(\frac{3i}{4} \partial_v \Delta_1^* + \frac{1}{4}\partial_v \Delta_2^*\right),  \nonumber \\
&\bar\psi_{b2}^{(1)(1)(0)} = \frac{\log2}{4\pi} \partial_v(\Delta_1^* - \Delta_2^*) + \left(\frac{3i}{4} \partial_v \Delta_2^* + \frac{1}{4}\partial_v \Delta_1^*\right).
\end{align}
Therefore, the first order time-derivative correction is
\begin{align}
\mathcal L_{eff}^{(1)(2)(0)} =& \frac{\log2}{8\pi} \left[(\Delta_2^* - \Delta_1^*)\partial_v (\Delta_2 - \Delta_1) - (\Delta_2 - \Delta_1)\partial_v (\Delta_2^* - \Delta_1^*) \right] \nonumber \\
& + \frac{3i}{8} \left[ (\Delta_2 \partial_v \Delta_2^* - \Delta_1^* \partial_v \Delta_1) -(\Delta_2^* \partial_v \Delta_2 - \Delta_1^* \partial_v \Delta_1)\right] \nonumber \\
& + \frac{1}{8} \left[(\Delta_2^* \partial_v \Delta_1 - \Delta_1^* \partial_v \Delta_2) + (\Delta_2 \partial_v \Delta_1^* - \Delta_1 \partial_v \Delta_2^*) \right] \nonumber \\
= & \frac{1}{8}  \left[ (1-3i) (\Delta_r \partial_v \Delta_a^* - \Delta_a^* \partial_v \Delta_r) + (1+3i) (\Delta_r^* \partial_v \Delta_a - \Delta_a \partial_v \Delta_r^*) \right] \nonumber \\
& + \frac{\log2}{8\pi} (\Delta_a^* \partial_v \Delta_a - \Delta_a \partial_v \Delta_a^*) \nonumber \\
= & -\frac{1}{4}(1-3i) \Delta_a^* \partial_v \Delta_r + \frac{1}{4}(1+3i) \Delta_r^* \partial_v \Delta_a + \frac{\log2}{4\pi} \Delta_a^* \partial_v \Delta_a, \label{Leff_120}
\end{align}
where in the last equality we dropped total time-derivative terms.

$\bullet$  $\mathcal L_{eff}^{(0)(4)(0)}$

We will be limited to quartic term $S_{eff}^{(0)(4)(0)}$, which ignores corrections from the $\delta\mu$ and time derivatives. Instead of directly using the formula \eqref{S_tot}, we play with the trick of \eqref{cancellation_proof} and simplify \eqref{S_tot} further. The quartic term in $\Delta_s$ of the bulk action is,
\begin{align}
S_0^{(0)(4)(0)} = & \int d^4x \int_{\infty_2}^{\infty_1} dr \sqrt{-g}\left[ - \frac{1}{2} F^{(0)(0)(0)} \cdot F^{(0)(4)(0)} -\frac{1}{4} \left(F^{(0)(2)(0)}\right)^2 \right] \nonumber \\
+ & \int d^4x \left[ - \frac{1}{2} r^5f(r)\Psi^{*(0)(3)(0)} \partial_r\Psi^{(0)(1)(0)}  - \frac{1}{2} r^5f(r)\Psi^{*(0)(1)(0)} \partial_r\Psi^{(0)(3)(0)} \right. \nonumber \\
&\left.- \frac{1}{2} r^5f(r)\Psi^{(0)(3)(0)} \partial_r\Psi^{*(0)(1)(0)} - \frac{1}{2} r^5f(r)\Psi^{(0)(1)(0)} \partial_r\Psi^{*(0)(3)(0)} \right]\bigg|_{\infty_2}^{\infty_1},
\end{align}
where we have utilized the scalar's EOMs only.
Note that $\sqrt{-g} F^{MN(0)(0)(0)}$ is constant in $r$. Consequently, regarding the gauge field's kinetic term, only $-\frac{1}{4} \left(F^{(0)(2)(0)}\right)^2$ will contribute. Integrating $\left(F^{(0)(2)(0)}\right)^2$ by parts and imposing the dynamical EOM for $A_v^{(0)(2)(0)}$, the quartic term of boundary effective Lagrangian is
\begin{align}
\mathcal L_{eff}^{(0)(4)(0)} = & \frac{1}{2} \left[ - \psi_{b1}^{(0)(3)(0)} \Delta_1^* + \psi_{b2}^{(0)(3)(0)} \Delta_2^* -\bar\psi_{b1}^{(0)(3)(0)} \Delta_1 + \bar \psi_{b2}^{(0)(3)(0)} \Delta_2  \right] \nonumber \\
& - \int_{\infty_2}^{\infty_1} dr \frac{rq^2}{f(r)} \bar A_v(r) A_v^{(0)(2)(0)}(r) \Psi^{*(0)(1)(0)}(r) \Psi^{(0)(1)(0)}(r). \label{Leff_040_v1}
\end{align}

First, we consider the scalar's surface term (i.e., the first line of \eqref{Leff_040_v1}), which requires to compute
\begin{align}
& \psi_{b1}^{(0)(3)(0)} = \frac{1}{2i\pi} \int_{\infty_2}^{\infty_1} dr u_2(r) \frac{-2q^2r}{f(r)} \bar A_v(r) A_v^{(0)(2)(0)}(r) \Psi^{(0)(1)(0)}(r), \nonumber \\
& \psi_{b2}^{(0)(3)(0)} = \frac{1}{2i\pi} \int_{\infty_2}^{\infty_1} dr u_1(r) \frac{-2q^2r}{f(r)} \bar A_v(r) A_v^{(0)(2)(0)}(r) \Psi^{(0)(1)(0)}(r), \nonumber \\
& \bar\psi_{b1}^{(0)(3)(0)} = \frac{1}{2i\pi} \int_{\infty_2}^{\infty_1} dr u_2(r) \frac{-2q^2r}{f(r)} \bar A_v(r) A_v^{(0)(2)(0)}(r) \Psi^{*(0)(1)(0)}(r), \nonumber \\
& \bar\psi_{b2}^{(0)(3)(0)} = \frac{1}{2i\pi} \int_{\infty_2}^{\infty_1} dr u_1(r) \frac{-2q^2r}{f(r)} \bar A_v(r) A_v^{(0)(2)(0)}(r) \Psi^{*(0)(1)(0)}(r), \label{psi_030}
\end{align}
where the explicit forms \eqref{jPsi_030} for source terms $j_\Psi^{(0)(3)(0)}$ and $j_{\Psi^*}^{(0)(3)(0)}$ have been plugged into \eqref{psi_b}. Here, the contour integral will be computed as in \eqref{contour_split}, and the contribution from the infinitesimal circle will vanish once $\epsilon \to 0$. Taking $\psi_{b1}^{(0)(3)(0)}$ as an example, it can be schematically rewritten as,
\begin{align}
\psi_{b1}^{(0)(3)(0)} = \frac{1}{2i\pi} \int_1^{\infty} dr \int_{\infty}^r dr^\prime \int_{\infty}^{r^\prime} dr^{\prime \prime} f_3(r, r^\prime, r^{\prime\prime}) + \frac{1}{2i \pi} \int_1^\infty dr f_4(r),
\end{align}
where
\begin{align}
f_3(r,r^\prime, r^{\prime\prime})=& \left\{u_2(r) \frac{-2q^2r}{f(r)} \bar A_v(r) \Psi^{(0)(1)(0)}(r) \frac{1}{r^{\prime 3}} j_v^{(0)(2)(0)}(r^{\prime\prime}) \right\} \bigg|_{r,r^\prime,r^{\prime\prime} \in [r_h, \infty_1)} \nonumber \\
&- \left\{u_2(r) \frac{-2q^2r}{f(r)} \bar A_v(r) \Psi^{(0)(1)(0)}(r) \frac{1}{r^{\prime 3}} j_v^{(0)(2)(0)}(r^{\prime\prime}) \right\}\bigg|_{r,r^\prime,r^{\prime\prime} \in [r_h, \infty_2)}, \nonumber \\
f_4(r) = &\left\{ u_2(r) \frac{-2q^2r}{f(r)} \bar A_v(r) \Psi^{(0)(1)(0)}(r) \frac{-c_1^{(0)(2)(0)}}{2r^2} \right\}\bigg|_{r\in[r_h, \infty_1)} \nonumber \\
& - \left\{ u_2(r) \frac{-2q^2r}{f(r)} \bar A_v(r) \Psi^{(0)(1)(0)}(r) \frac{-c_2^{(0)(2)(0)}}{2r^2} \right\}\bigg|_{r\in[r_h, \infty_2)}.
\end{align}
Here, the integration constants $c_1^{(0)(2)(0)}$ and $c_2^{(0)(2)(0)}$ are defined in \eqref{c12lmn}, and are of quadratic order in the order parameter. The numerical results for the integration constants $c_1^{(0)(2)(0)}$ and $c_2^{(0)(2)(0)}$ are
\begin{align}
c_1^{(0)(2)(0)}=& (0.483498 - 0.110318 i)\Delta_1 \Delta_1^* + (0.0165017 + 0.0551589 i) (\Delta_1 \Delta_2^* + \Delta_2 \Delta_1^*) \nonumber \\
& -0.0165017 \Delta_2 \Delta_2^*, \nonumber \\
c_2^{(0)(2)(0)}=& -0.0165017 \Delta_1 \Delta_1^* + (0.0165017 - 0.0551589 i) (\Delta_1 \Delta_2^* + \Delta_2 \Delta_1^*) \nonumber \\
& + (0.483498 + 0.110318 i) \Delta_2 \Delta_2^*.
\end{align}
The numerical results for $\psi_b$'s are
\begin{align}
2i \pi \psi_{b1}^{(0)(3)(0)}=& (0.00658623 + 0.124096 i) \Delta_1 \Delta_1 \Delta_1^* -(0.00162114 - 0.00340208 i) \Delta_1 \Delta_1 \Delta_2^*  \nonumber \\
&-(0.00324227 - 0.00680415 i) \Delta_1 \Delta_2 \Delta_1^* -0.0000677893\Delta_1 \Delta_2 \Delta_2^*  \nonumber \\
& -0.0000338947 \Delta_2 \Delta_2 \Delta_1^*  -(0.00162114 + 0.00340208 i) \Delta_2 \Delta_2 \Delta_2^* , \nonumber \\
2i \pi \psi_{b2}^{(0)(3)(0)}=& (0.00162114 - 0.00340208 i) \Delta_1 \Delta_1 \Delta_1^* + 0.0000338947 \Delta_1 \Delta_1 \Delta_2^*  \nonumber \\
&+ 0.0000677893 \Delta_1 \Delta_2 \Delta_1^* + (0.00324227 + 0.00680415 i) \Delta_1 \Delta_2 \Delta_2^*  \nonumber \\
& + (0.00162114 + 0.00340208 i)\Delta_2 \Delta_2 \Delta_1^*  - (0.00658623 - 0.124096 i) \Delta_2 \Delta_2 \Delta_2^* , \nonumber \\
2i \pi \bar \psi_{b1}^{(0)(3)(0)}=& (0.00658623 + 0.124096 i) \Delta_1 \Delta_1^* \Delta_1^* -(0.00162114 - 0.00340208 i) \Delta_2 \Delta_1^* \Delta_1^*  \nonumber \\
&-(0.00324227 - 0.00680415 i) \Delta_1 \Delta_1^* \Delta_2^* -0.0000677893 \Delta_2 \Delta_1^* \Delta_2^*  \nonumber \\
& -0.0000338947 \Delta_1 \Delta_2^* \Delta_2^*  -(0.00162114 + 0.00340208 i) \Delta_2 \Delta_2^* \Delta_2^* , \nonumber \\
2i \pi \bar\psi_{b2}^{(0)(3)(0)}=& (0.00162114 - 0.00340208 i) \Delta_1 \Delta_1^* \Delta_1^* + 0.0000338947 \Delta_2 \Delta_1^* \Delta_1^*  \nonumber \\
&+ 0.0000677893  \Delta_1 \Delta_1^* \Delta_2^* + (0.00324227 + 0.00680415 i) \Delta_2 \Delta_1^* \Delta_2^*  \nonumber \\
& + (0.00162114 + 0.00340208 i) \Delta_1 \Delta_2^* \Delta_2^*  -(0.00658623 - 0.124096 i) \Delta_2 \Delta_2^* \Delta_2^*.
\end{align}
Again $2i \pi \psi_{bs}^{(0)(3)(0)}$ and $2i \pi \bar{\psi}_{bs}^{(0)(3)(0)}$ are related by charge conjugation: $\Delta_s\leftrightarrow{\Delta}^*_s$.
In terms of $(r,a)$-basis, the scalar's surface term is
\begin{align}
&\frac{1}{2} \left[ - \psi_{b1}^{(0)(3)(0)} \Delta_1^* + \psi_{b2}^{(0)(3)(0)} \Delta_2^* -\bar\psi_{b1}^{(0)(3)(0)} \Delta_1 + \bar \psi_{b2}^{(0)(3)(0)} \Delta_2  \right] \nonumber \\
%=& 0.000258012 i (\Delta_a \Delta_a^*)^2 -0.00933375  \Delta_a \Delta_a^* \Delta_a^* \Delta_r + 0.000526813 i (\Delta_a^* \Delta_r)^2 \nonumber \\
%& -0.00933375  \Delta_a^2 \Delta_a^* \Delta_r^* + 0.00210725 i\Delta_a \Delta_r \Delta_a^* \Delta_r^* -0.0416667 \Delta_a^* \Delta_r^* \Delta_r^2 \nonumber \\
%& + 0.000526813 i (\Delta_a \Delta_r^*)^2 -0.0416667  \Delta_a \Delta_r \Delta_r^{*2} \nonumber \\
=& 0.000258012 i (\Delta_a \Delta_a^*)^2 -0.00933375  \Delta_a \Delta_a^*( \Delta_a^* \Delta_r + \Delta_a \Delta_r^*) \nonumber \\
& + 0.000526813 i\left[ (\Delta_a^* \Delta_r)^2 + (\Delta_a \Delta_r^*)^2 \right] + 0.00210725 i\Delta_a \Delta_r \Delta_a^* \Delta_r^* \nonumber \\
&-0.0416667 (\Delta_a^* \Delta_r^* \Delta_r^2 + \Delta_a \Delta_r \Delta_r^{*2} ). \label{scalar_surface_040}
\end{align}

Next, we compute the contribution from the gauge field's kinetic term:
\begin{align}
I_5=- \int_{\infty_2}^{\infty_1} dr \frac{rq^2}{f(r)} \bar A_v(r) A_v^{(0)(2)(0)}(r) \Psi^{*(0)(1)(0)}(r) \Psi^{(0)(1)(0)}(r),
\end{align}
which will be calculated in the same way of $\psi_{b1}^{(0)(3)(0)}$. By splitting the radial contour as in \eqref{contour_split}, the bulk term $I_5$ is reduced into a radial integral on a single Schwarzschild-AdS$_5$ space:
\begin{align}
I_5= \int_1^{\infty} dr \int_{\infty}^r dr^\prime \int_{\infty}^{r^\prime} dr^{\prime\prime} f_5(r,r^\prime, r^{\prime\prime}) + \int_1^{\infty} dr f_6(r),
\end{align}
where
\begin{align}
f_5(r,r^\prime, r^{\prime\prime})= & \left\{\frac{rq^2}{f(r)} \bar A_v(r) \Psi^{*(0)(1)(0)}(r) \Psi^{(0)(1)(0)}(r) \frac{1}{r^{\prime3}} j_v^{(0)(2)(0)}(r^{\prime\prime}) \right\} \bigg|_{r,r^\prime, r^{\prime\prime} \in [r_h, \infty_2)} \nonumber \\
&-\left\{\frac{rq^2}{f(r)} \bar A_v(r) \Psi^{*(0)(1)(0)}(r) \Psi^{(0)(1)(0)}(r) \frac{1}{r^{\prime3}} j_v^{(0)(2)(0)}(r^{\prime\prime}) \right\} \bigg|_{r,r^\prime, r^{\prime\prime} \in [r_h, \infty_1)}, \nonumber \\
f_6(r) = & \left\{\frac{rq^2}{f(r)} \bar A_v(r) \Psi^{*(0)(1)(0)}(r) \Psi^{(0)(1)(0)}(r) \frac{-c_2^{(0)(2)(0)}}{2r^2} \right\} \bigg|_{r\in [r_h, \infty_2)} \nonumber \\
& - \left\{\frac{rq^2}{f(r)} \bar A_v(r) \Psi^{*(0)(1)(0)}(r) \Psi^{(0)(1)(0)}(r) \frac{-c_1^{(0)(2)(0)}}{2r^2} \right\} \bigg|_{r\in [r_h, \infty_1)}.
\end{align}
The numerical result for $I_5$ is
\begin{align}
I_5=& - 0.000129006 i (\Delta_a \Delta_a^*)^2 +0.00466688  \Delta_a \Delta_a^*( \Delta_a^* \Delta_r + \Delta_a \Delta_r^*) \nonumber \\
& - 0.000263406 i \left[ (\Delta_a^* \Delta_r)^2 + (\Delta_a \Delta_r^*)^2 \right] - 0.00105363 i\Delta_a \Delta_r \Delta_a^* \Delta_r^* \nonumber \\
&+ 0.0208333  (\Delta_a^* \Delta_r^* \Delta_r^2 + \Delta_a \Delta_r \Delta_r^{*2} ),
\end{align}
which is minus half of the scalar's surface term \eqref{scalar_surface_040}.

Eventually, the quartic order effective Lagrangian is
%\begin{align}
%\mathcal L_{eff}^{(0)(4)(0)} = & 0.000129006 i (\Delta_a \Delta_a^*)^2 - 0.00466688  \Delta_a \Delta_a^*( \Delta_a^* \Delta_r + \Delta_a \Delta_r^*) \nonumber \\
%& + 0.000263406 i \left[ (\Delta_a^* \Delta_r)^2 + (\Delta_a \Delta_r^*)^2 \right] + 0.00105363 i\Delta_a \Delta_r \Delta_a^* \Delta_r^* \nonumber \\
%&- 0.0208333  (\Delta_a^* \Delta_r^* \Delta_r^2 + \Delta_a \Delta_r \Delta_r^{*2} ), \label{Leff_040}
%\end{align}
\begin{align}
\mathcal L_{eff}^{(0)(4)(0)} = & a_1 i (\Delta_a \Delta_a^*)^2 +a_2  \Delta_a \Delta_a^*( \Delta_a^* \Delta_r + \Delta_a \Delta_r^*) + a_3 i \left[ (\Delta_a^* \Delta_r)^2 + (\Delta_a \Delta_r^*)^2 \right] \nonumber \\
& + a_4 i \Delta_a \Delta_r \Delta_a^* \Delta_r^* + a_5  (\Delta_a^* \Delta_r^* \Delta_r^2 + \Delta_a \Delta_r \Delta_r^{*2} ), \label{Leff_040}
\end{align}
where various coefficients are
\begin{align}
&a_1= 0.000129006 , \qquad a_2 = -0.00466688, \qquad a_3 = 0.000263406, \nonumber \\
&a_4 = 0.001053630  , \qquad a_5= -0.02083330.
\end{align}
Indeed, for the path integral based on effective action to be well-defined, the coefficients $a_1, a_3, a_4$ must be non-negative.
\eqref{Leff_020}, \eqref{Leff_021}, \eqref{Leff_120} and \eqref{Leff_040} are the main results of the paper.

{\bf Check various constraints}

$\bullet$ $U(1)$ symmetry

Since we are in the high temperature phase, the global $U(1)$ symmetry on the boundary is preserved. For a non-equilibrium EFT, the $U(1)_1\times U(1)_2$ symmetry associated with SK contour is reduced into the diagonal one \cite{Crossley:2015evo,Glorioso:2018mmw}:
\begin{align}
\Delta_s \to e^{iq \Lambda} \Delta_s, \qquad \Delta_s^* \to e^{-iq \Lambda} \Delta_s^*, \qquad \mu \to \mu, \qquad s=1,2. \label{U(1)diagonal}
\end{align}
where the transformation parameter $\Lambda$ is a constant for our situation. The diagonal symmetry \eqref{U(1)diagonal} is perfectly satisfied by our results: the phase factors cancel among equal number of $\Delta$ and $\Delta^*$ in each term.

%
%In the appendix \ref{boundary_action_derivation} we have demonstrated that the bulk gauge invariance is preserved in the partially on-shell formalism. That is, thus-derived boundary action is invariant under the following $U(1)_1\times U(1)_2$ transformations:
%\begin{align}
%\Delta_s \to e^{iq \Lambda_s}\Delta_s, \quad \Delta_s^* \to e^{-iq \Lambda_s}\Delta_s^*, \quad \mathcal A_\mu \to \mathcal A_\mu - \partial_\mu \Lambda_s, \quad \varphi_s \to \varphi_s + \Lambda_s, \quad s=1,2, \label{U(1)12}
%\end{align}
%where $\Lambda_s= \Lambda_2(v,\vec x)$.
%However, thanks to special limitations made for the chemical potential $\mu$, the symmetry \eqref{U(1)12}

$\bullet$ $Z_2$-reflection symmetry

Under the exchange of $\Delta_1 \leftrightarrow \Delta_2$ and $\Delta^*_1 \leftrightarrow \Delta_2^*$, the effective Lagrangian $\mathcal L_{eff}[\Delta_1, \Delta_1^*; \Delta_2, \Delta_2^*]$ obeys \cite{Crossley:2015evo}
\begin{align}
\left(\mathcal L_{eff}[\Delta_1, \Delta_1^*; \Delta_2, \Delta_2^*]\right)^* = - \mathcal L_{eff}[\Delta_2, \Delta_2^*; \Delta_1, \Delta_1^*]. \label{Z2_symmetry}
\end{align}
The $Z_2$ symmetry is actually a consistency condition. In \cite{Crossley:2015evo} the $Z_2$ symmetry \eqref{Z2_symmetry} was employed to constrain the effective action for real scalar fields: the term containing even numbers of $a$-variables must be purely imaginary, and the term containing odd numbers of $a$-variables should be purely real. With complex fields in our case, the coefficients can be complex in general. Nevertheless, \eqref{Z2_symmetry} remains true for our results. Particularly, terms with complex coefficients in \eqref{Leff_120} map to each other upon integration by parts.
%Indeed, the $U(1)$ symmetry \eqref{U(1)diagonal} and the $Z_2$ symmetry \eqref{Z2_symmetry} can be used to fix the most general structures for $\mathcal L_{eff}$.
%{\color{red}I checked \eqref{Z2_symmetry} holds for all other $\mathcal L_{eff}$ except $\mathcal L_{eff}^{(1)(2)(0)}$, for which the sign should be flipped. The conclusion applies to both holographic and weakly coupled results.
%  I also feel the $Z_2$ is some equivalence of time-reversal symmetry. Both are anti-linear operator sending $i\to-i$, $\Delta_s\to\Delta^*_s$ and $\partial_v^* \to - \partial_v$. So we may also say all other $\mathcal L_{eff}$ except $\mathcal L_{eff}^{(1)(2)(0)}$ are T-odd. This is very puzzling to me as our bulk Lagrangian is T-even.}

$\bullet$ Kubo-Martin-Schwinger (KMS) symmetry

Recall that in the $\delta\mu$-corrections and quartic order terms, we do not cover the first order time-derivative terms. Thus, the cross-check of KMS constraints will be limited to $\mathcal L_{eff}^{(0)(2)(0)}$ and the $ra$-terms in $\mathcal L_{eff}^{(1)(2)(0)}$ \cite{Chou:1984es,Wang:1998wg,Carrington:2000zn,Crossley:2015evo}. Note that $\Delta_{r/a}$ couple to $\psi_{b,{a/r}}$ in the Lagrangian, we may relate the coefficients to the correlator of $\psi_{b}$ as \cite{Son:2009vu}
  \begin{align}\label{psib_G}
    G^S(x,y)=\langle\psi_{b,r}(x)\bar{\psi}_{b,r}(y)\rangle&=\frac{i\delta^2 S_{eff}}{\delta\Delta_a(x)\delta\Delta^*_a(y)}, \nonumber\\
    iG^R(x,y)=\langle\psi_{b,r}(x)\bar{\psi}_{b,a}(y)\rangle&=\frac{i\delta^2 S_{eff}}{\delta\Delta_a(x)\delta\Delta^*_r(y)}, \nonumber\\
    iG^A(x,y)=\langle\psi_{b,a}(x)\bar{\psi}_{b,r}(y)\rangle&=\frac{i\delta^2 S_{eff}}{\delta\Delta_r(x)\delta\Delta^*_a(y)}.
  \end{align}
%\MF{What is the definition of $S_{eff}$? I find that $G^R=\delta \mu -(1+3 i)i\omega /4$ from eq. (4.22). }
The symmetric and retarded two-point correlators read
\begin{align}
G^S = \frac{1}{2\pi} - \delta\mu \frac{\log2}{\pi} + \mathcal{O}(\omega), \qquad G^R= \delta \mu - \frac{1}{4}(1+3i)i\omega + \mathcal{O}(\omega^2), \label{G_sym_ret}
\end{align}
which obeys the standard fluctuation-dissipation relation $G^S = -\coth[\omega/(2T)] {\rm Im}(G^R)$ up to $\delta\mu$-term at lowest order in $\omega$\footnote{Discussions on the fluctuation-dissipation relation for correlators among different fields can be found in Section 2.3 of \cite{Kovtun:2012rj}.}.

$\bullet$ (Generalized) Onsager relation

Just as $Z_2$ symmetry, the (generalized) Onsager relation is also a consistency condition \cite{Crossley:2015evo}. The advanced two-point correlator reads
\begin{align}
G^A= \delta\mu + \frac{1}{4} (1-3i) i\omega + \mathcal{O}(\omega^2), \label{G_adv}
\end{align}
which satisfies the familiar Onsager relation $\left(G^R\right)^*=G^A$.

The generalized Onsager relations involve higher-point correlators with one $r$-index only. For our case, the only relevant components are
\begin{align}\label{four-point}
  iG_{raaa}(x,y,z,w)&=\langle\psi_{b,r}(x)\bar{\psi}_{b,a}(y)\psi_{b,a}(z)\bar{\psi}_{b,a}
  (w)\rangle= \frac{1}{2}\frac{i\delta^4 S_{eff}}{\delta\Delta_a(x)\delta\Delta^*_r(y) \delta\Delta_r(z)\delta\Delta^*_r(w)}, \nonumber\\
  iG_{araa}(x,y,z,w)&=\langle\psi_{b,a}(x)\bar{\psi}_{b,r}(y)\psi_{b,a}(z)\bar{\psi}_{b,a}
  (w)\rangle=\frac{1}{2}\frac{i\delta^4 S_{eff}}{\delta\Delta_r(x)\delta\Delta^*_a(y)
  \delta\Delta_r(z)\delta\Delta^*_r(w)}.
\end{align}
They characterize the response of ${\psi}_b$ and $\bar{\psi}_b$ to external sources cubic in $\Delta$ and $\Delta^*$. Our results indicate $G_{raaa}=G_{araa}$ and are real. This may be viewed as analog of $(G^R)^*=G^A$ to the lowest order in $\omega$\footnote{This is similar to (2.60) of \cite{Crossley:2015evo} except the $U(1)$ symmetry discussed above excludes possible correlators with unequal number of $\psi_b$ and $\bar{\psi}_b$.}.

%The generalized Onsager relations involve higher-point correlators with one $r$-index only, like $G_{raa\cdots aa}$. To check this relation based on our quartic terms $\mathcal L_{eff}^{(0)(4)(0)}$, we need the effective action with sources (coupled to $\Delta_s$) turned on. The $U(1)$ symmetry inspires to obtain sourced action from $S_{eff}[\Delta_s,\Delta_s^*]$ by making the following replacement
%\begin{align}
%\Delta_s \to \Delta_s + \phi_s, \qquad \Delta_s^* \to \Delta_s^* + \phi_s^*, \qquad s=1,2.
%\end{align}
%Then, the source action (with dynamical variables switched off) is simply $S_{eff}[\Delta_s \to \phi_s, \Delta_s^* \to \phi_s^*]$, from which we directly obtain $G_{raaa}$ etc. Indeed, from the last term of \eqref{Leff_040}, it is obvious that all four-point correlators with only one $r$-index are identical and purely real, in agreement with (2.60) of \cite{Crossley:2015evo}. Here, we see this consistency condition is implied/guaranteed by the $U(1)$ symmetry or the $Z_2$ symmetry.

{\bf Comparison with weakly coupled results \cite{Levchenko2007,Kamenev2011}.}

The relevant results to quote are (14.70)-(14.72b) of \cite{Kamenev2011}:
\begin{align}
S_{\rm GL}= 2\nu {\rm Tr} \left\{ \vec \Delta_{\mathcal{K}}^{\dag}(\vec r,t) {\hat L}^{-1}  \vec \Delta_{\mathcal{K}}(\vec r,t) \right\},
\end{align}
where $\vec\Delta_{\mathcal K}=(\vec\Delta_{\mathcal K}^{\rm cl}, \vec\Delta_{\mathcal K}^{\rm q})^{\rm T}$ denotes the classical part and fluctuation of the order parameter (the scalar condensate). The matrix ${\hat L}^{-1}$ is given by
\begin{equation}
{\hat L}^{-1}=
\left(
\begin{array}{cc}
0 & (L^{-1})^A\\
(L^{-1})^R & (L^{-1})^K \\
\end{array}
\right), \label{h_c123}
\end{equation}
with matrix elements given as
\begin{align}
&(L^{-1})^{R(A)}= \frac{\pi}{8T}\left[ \mp \partial_t +D (\nabla_r + 2ie A_\mathcal{K}^{\rm cl})^2 -\tau_{\rm GL}^{-1} - \frac{7\zeta(3)}{\pi^3 T_c} |\Delta_\mathcal{K}^{\rm cl}|^2  \right], \nonumber \\
&(L^{-1})^K= \coth\frac{\omega}{2T} \left[ (L^{-1})^R(\omega) - (L^{-1})^A(\omega) \right] \approx \frac{i\pi}{2}, \label{L_KRA}
\end{align}
where $R$ ($A$) corresponds to $-$ ($+$) in the time-derivative term and $\tau_{\rm GL}=\pi/[8(T-T_c)]$. The subscript ``$\mathcal K$'' denotes the so-called $\mathcal K$-gauge, in which the time-component of external gauge potential is set to zero. Apparently, the second equality of \eqref{L_KRA} comes from KMS conditions.
Explicitly, the action $S_{\rm GL}$ is
\begin{align}
S_{\rm GL}= 2\nu\int d^4x \left[ \Delta_\mathcal{K}^{\rm q*} (L^{-1})^R \Delta_\mathcal{K}^{\rm cl} + \Delta_\mathcal{K}^{\rm cl*} (L^{-1})^A \Delta_\mathcal{K}^{\rm q} + \Delta_\mathcal{K}^{\rm q*} (L^{-1})^K \Delta_\mathcal{K}^{\rm q} \right] \label{S_GL}
\end{align}
In the action $S_{\rm GL}$, there are two ``fundamental parameters'': $\nu$---the density of states, and $D$---the diffusion constant.

Since we have no gauge potentials but only constant chemical potential, we may make the identification $(\Delta_{\mathcal K}^{\rm cl}, \Delta_{\mathcal K}^{\rm q})^{\rm T}= (\Delta_r, \Delta_a)^{\rm T}$ and rewrite the weakly coupled results in our notations as
\begin{align}\label{weak_L}
  \mathcal L_{eff}^{(0)(2)(0)}&=i\pi\nu\Delta^*_a\Delta_a, \nonumber\\
  \mathcal L_{eff}^{(0)(2)(1)}&=\frac{\pi\nu}{4T}(-\tau_{\rm GL}^{-1})(\Delta_a\Delta^*_r+\Delta^*_a\Delta_r), \nonumber\\
  \mathcal L_{eff}^{(1)(2)(0)}&=\frac{\pi\nu}{4T}(-\Delta^*_a\partial_v\Delta_r+\Delta^*_r
  \partial_v\Delta_a), \nonumber\\
  \mathcal L_{eff}^{(0)(4)(0)}&=\frac{\pi\nu}{4T}\left(-\frac{7\zeta(3)}{\pi^3T_c}\right)
  |\Delta_r|^2(\Delta_a\Delta^*_r+\Delta^*_a\Delta_r).
\end{align}
where the spatial derivative terms are ignored and $A_{\mathcal K}^{\rm cl}$ is set to zero. 
%\sout{{ The quadratic terms $\mathcal L_{eff}^{(0)(2)(1)}$ and $\mathcal L_{eff}^{(1)(2)(0)}$ predict the critical slowing down of the condensate, which is a non-equilibrium characteristic of the study.}}
Despite the difference in model details, the holographic and weakly coupled results do share some similarities:

  $\bullet$ The results of $\mathcal L_{eff}^{(0)(2)(0)}$ have the same structure with purely imaginary coefficients.

  $\bullet$ The results of $\mathcal L_{eff}^{(0)(2)(1)}$ contain a common term $\Delta_a\Delta^*_r+\Delta^*_a\Delta_r$, whose coefficients scale the same near the critical point: $\delta\mu=\mu-\mu_0\sim T_c-T$ and $\tau_{\rm GL}^{-1}\sim T-T_c$. It can be understood from universality. Both models belong to model A of Hohenberg-Halperin classification for dynamic unversality class, with the following static and dynamic critical exponents $\nu=1/2$ and $z=2$, which dicates the scaling of relaxation time $\tau_{\rm GL}\sim (T-T_c)^{-z\nu}\sim (T-T_c)^{-1}$, see \cite{Maeda:2009wv,Bu:2019epc} for calculations of crtical exponents for s-wave and p-wave holographic superconductors.
{The quadratic terms $\mathcal L_{eff}^{(0)(2)(1)}$ and $\mathcal L_{eff}^{(1)(2)(0)}$ predict the critical slowing down of the condensate, which is a non-equilibrium characteristic of the effective action.} 

Now, let us see the main differences between holographic and weakly coupled results:% our results and those of \cite{Kamenev2011,Levchenko2007}:

$\bullet$  The coefficients of $\Delta^*_a\partial_v\Delta_r$ and $\Delta^*_r\partial_v\Delta_a$ are complex in \eqref{Leff_120}, while they are purely real in \eqref{weak_L}.
In momentum space, the holographic results \eqref{G_sym_ret} and \eqref{G_adv}
%
%
%  give
%  \begin{align}\label{psib_G}
%    iG^R(\omega)=-\frac{1}{4}(1-3i)\omega, \nonumber\\
%    iG^A(\omega)=\frac{1}{4}(1+3i)\omega,
%  \end{align}
indicate the retarded and advanced correlators of $\psi_{b}$ contain also real parts, which are not constrained by KMS condition. This is also consistent with the $Z_2$ symmetry.

%{it might be too early to mention $Z_2$ as we only introduce it below}\sout{However, from both the $Z_2$-symmetry and the local KMS conditions, it is not excluded that the coefficients can contain imaginary parts.}

$\bullet$ The $\Delta_a^* \Delta_a$-term in $\mathcal L_{eff}^{(0)(2)(1)}$ is absent in \cite{Kamenev2011,Levchenko2007}. Note this term is of the same structure as the lowest order result \eqref{Leff_020}, for which the chemical potential is set to the critical value $\mu_0$. While the free energy is a minimum at the critical point, the fluctuation of order parameter is not necessarily an extremum. Indeed, $\mathcal L_{eff}^{(0)(2)(1)}$ corresponds to the correction to the order parameter fluctuation away from the critical point.
%Generically away from the critical point, we would expect that such a term will have a coefficient $r_h^2f_0(\mu^2/T^2)$, which covers relevant results in \eqref{Leff_020} and \eqref{Leff_021}.
Additionally, by KMS condition, the coefficient of $\Delta_a^* \Delta_a$-term in $\mathcal L_{eff}^{(0)(2)(1)}$ shall be related to that of first order time-derivative correction to $\delta \mu (\Delta_a \Delta_r^* + \Delta_a^* \Delta_r)$.

$\bullet$ The time-derivative term $\Delta_a^* \partial_v \Delta_a - \Delta_a \partial_v \Delta_a^*$ in $\mathcal L_{eff}^{(1)(2)(0)}$ was not covered in \cite{Kamenev2011,Levchenko2007} and shall be constrained by KMS conditions. However, this requires second order time-derivative corrections to $ra$-terms, which are beyond the scope of present work.

$\bullet$ At the quartic order, only $arrr$-terms are covered in \cite{Kamenev2011,Levchenko2007}, while our results contain all possible structures. {These extra terms represent nonlinear interactions among noises, as well as nonlinear interactions between the order parameter and noise \cite{Crossley:2015evo}. Particularly, they cannot be accounted for in the time-dependent Ginzburg-Landau equation \cite{Kamenev2011,Levchenko2007}. It will be interesting to explore their physical consequences based on the EFT framework.}

\section{Summary and outlook} \label{summary}

We have initiated the construction of SK effective action for a holographic superconductor model in the spatially homogeneous limit. Near the critical point, we obtained the time-dependent Ginzburg-Landau effective action in the hydrodynamic limit. The effective action is accurate up to quartic order in the fluctuating scalar condensate. Compared to the results obtained for weakly coupled BCS superconductor \cite{Levchenko2007,Kamenev2011}, our effective action contains more structures allowed by general physical considerations. Our study demonstrates that the holographic prescription of \cite{Glorioso:2018mmw} for SK contour for non-equilibrium state works very well for nonlinear problems in the bulk.

%It is also interesting to go beyond first order in time-derivative expansion.

The present study is limited to spatially homogeneous case, which renders the charge diffusion part (i.e., the normal current part $S_{\rm N}$) to be decoupled. It is interesting to go beyond this approximation, and explore the interaction between the fluctuating charge density and the fluctuating order parameter, which is supposed to give rise to more interesting physics\cite{Levchenko2007,Kamenev2011}.

Another interesting further project will be to include backreaction of bulk gravity. At quadratic order, an extra diffusive mode (coming from the energy/momentum on the boundary) will mix with the charge diffusion, and a prorogating sound mode will emerge and will be coupled to the scalar condensate (or the Goldstone mode). Beyond the linear regime, we will see an interacting EFT for dissipative superfluid, which will be useful for investigating fluctuation/dissipation effects in phenomena such as scattering process between phonon and vortex \cite{Nicolis:2013lma}.

We hope to address these interesting questions in the near future.

%{\color{red} Some open questions from my side:}
%
%\noindent $\bullet$ Can we recover the Ginzburg-Landau free energy from the EFT action? If yes, how? The effective action can be used to derive the time-dependent Ginzburg-Landau equation (stochastic equation), as done in \cite{Kamenev2011,Levchenko2007}. {\color{blue}I think by dropping the dissipative $\partial_v$ terms in $(L^{-1})^{R/A}$ and fluctuating term of $(L^{-1})^K$.}
%
%
%\noindent $\bullet$  It was claimed that the derivation of \cite{Kamenev2011,Levchenko2007} is in the high temperature phase. Our derivation is on the critical point and then we deviate from it. Are we in the high temperature or low temperature phase? This might affect the extraction of Ginzburg-Landau relaxation time $\tau_{\rm GL}$ (particularly the sign).{We are in the high-temperature phase so that we only have fluctuation of condensate not a mean field vev.}
%
%
%\noindent $\bullet$ Our notation for external gauge potential is misleading. I think the $\mathcal A_\mu$ in the main text (e.g. in \eqref{asymp_expand_contour}) should be understood as the ``gauge invariant combination'' $\mathcal A_\mu + \partial_\mu \varphi$. Practically, we turned off the diffusive field $\varphi$. BTW, the chemical potential is identified with $\mathcal A_v + \partial_v \varphi$.
%
%\noindent $\bullet$ {I want to add one: if we Legendre transform the effective action to obtain the generating functional. Can we discuss cross-over of normal dynamics and critical slowing-down? I guess it might be possible through interplay of $\partial_v$ and $\delta\mu$ terms.}

\appendix

\section{Boundary effective action from bulk path integral} \label{boundary_action_derivation}

In this appendix, we provide further details on a formal derivation of boundary effective action from partially on-shell bulk solution \cite{Crossley:2015tka}. We will closely follow \cite{Crossley:2015tka,Bu:2020jfo}, based on early attempts \cite{Heemskerk:2010hk,Faulkner:2010jy,Nickel:2010pr}. We would like to demonstrate that once dynamical EOMs are correctly chosen (in compatible with the chosen gauge convention), the resultant partially on-shell bulk action takes the same form.

Consider the bulk path integral \eqref{Z_AdS}:
\begin{align}
Z_{\rm AdS}= \int D A_r^\prime DA_\mu^\prime D \Psi^\prime D\Psi^{*\prime} e^{iS_0[A_M^\prime, \Psi^\prime, \Psi^{*\prime}]+S_{\rm bdy}^{\rm II}},
\end{align}
where primed fields $A_M^\prime, \Psi^\prime, \Psi^{*\prime}$ denote bulk field configurations without any gauge-fixing.
In the saddle point approximation, the computation of $Z_{\rm AdS}$ boils down to solving the classical field equations in the bulk.
According to \cite{Glorioso:2018mmw,Crossley:2015tka}, integrating out $A_\mu^\prime, \Psi^\prime, \Psi^{*\prime}$ would yield boundary effective action:
\begin{align}
Z_{\rm AdS} & = \int D A_r^\prime \int D \Delta^\prime D \bar \Delta^\prime e^{iS_0[A_r^\prime, A_\mu^\prime[A_r^\prime], \Psi^\prime[A_r^\prime], \Psi^{*\prime}[A_r^\prime], \Delta^\prime, \bar \Delta^\prime]+S_{\rm bdy}^{\rm II}} \nonumber \\
&\simeq \int D\varphi \int D \Delta D \bar \Delta e^{i S_{eff}[\varphi, \Delta, \bar\Delta]}. \label{Z_AdS_pos}
\end{align}
Here, the fact that $S_0$ depends on $A_r^\prime$ indicates we go via the partially on-shell procedure, namely, leaving aside the constraint equation unsolved. In \eqref{Z_AdS_pos} the fluctuation of $\Delta, \bar \Delta$ is introduced by Dirichlet-like boundary conditions for $\Psi, \Psi^*$, i.e., under the second quantization scheme as implied by the usage of $S_{\rm bdy}^{\rm II}$.

In \eqref{Z_AdS_pos} we want to clarify two issues as going from the first equality to the second one. First, we explain why an integration over $A_r^\prime$ is equivalent to the integration over the dynamical field $\varphi$ (responsible for the charge diffusion), following \cite{Crossley:2015tka,Glorioso:2018mmw}. Second, we will make clear relationship between $S_{eff}$ and the bulk action $S_0$, say \eqref{S_eff_S0_pos}, and particularly show that such an identification based on the off-shell procedure is free of ambiguity.

The first issue could be understood with the help of bulk $U(1)$ gauge symmetry \cite{Nickel:2010pr}, as briefly discussed in section \ref{holo_setup}. We elaborate on the analysis here.
Consider a gauge transformation over $A_M^\prime, \Psi^\prime, \Psi^{*\prime}$:
\begin{align}
A_M^\prime \to A_M = A_M^\prime + \partial_M \Lambda, \qquad \Psi^\prime \to \Psi= \Psi^\prime e^{iq \Lambda}, \qquad \Psi^{*\prime} \to \Psi^*= \Psi^{*\prime} e^{-iq \Lambda}, \label{gauge_trans}
\end{align}
which can bring a specific field configuration to be in an any gauge. On the boundary, \eqref{gauge_trans} reads as
\begin{align}
\mathcal A_\mu^\prime \to \mathcal A_\mu = \mathcal A_\mu^\prime + \partial_\mu \varphi, \qquad \Delta^\prime \to \Delta= \Delta^\prime e^{iq \varphi}, \qquad \bar\Delta^\prime \to \bar\Delta = \bar\Delta^\prime e^{-iq \varphi},
\end{align}
where $\varphi(x^\alpha) \equiv \Lambda(r=\infty,x^\alpha)$ is identified with the dynamical variable for the charge diffusion. Therefore, going from field configuration $(A_M^\prime, \Psi^\prime, \Psi^{*\prime})$ without gauge-fixing to the gauge-fixed one $(A_r=0, A_\mu, \Psi, \Psi^*)$ or $(A_r =- A_v/(r^2f(r)), A_\mu, \Psi, \Psi^*)$ is equivalent to changing integration variable from $A_r^\prime$ to $\Lambda$ (or, equivalently, $\varphi$). Here, we are not rigorous about the potetial Jacobian determinant due to change of integration variables, as it will not affect calculations of physical observables.

Now we turn to the second issue. With the integration over $A_r^\prime$ identified with the integration over $\varphi$, it is straightforward to conclude the relationship \eqref{S_eff_S0_pos}: the boundary effective action is just identified with the renormalized partially on-shell bulk action. Now we want to clarity such a procedure does not depend on gauge choice and is thus free of ambiguity. The following gauge invariant objects will be useful for later derivation:
\begin{align}
F^{\prime MN} = F^{MN}, \qquad  \Psi^{*\prime} D^{\prime M} \Psi^\prime = \Psi^* D^M \Psi, \qquad \Psi^\prime (D^M \Psi)^{*\prime} = \Psi (D^M \Psi)^*. \label{gauge_invariant}
\end{align}
Note that the action $S_0$ in \eqref{Z_AdS_pos} is computed with partially on-shell bulk solution without any gauge-fixing, which is not easy implement in practice. Now, we will compute the partially on-shell bulk action $S_0$ with the help of gauge transformation \eqref{gauge_trans}. The primed version of \eqref{S0_pos_Schw_radial} is computed as
\begin{align}
S_0 = & -\int d^4x \sqrt{-\gamma} n_M \left( \frac{1}{2} A_N^\prime F^{\prime MN} + \frac{1}{2} \Psi^{*\prime} D^{\prime M} \Psi^\prime  + \frac{1}{2} \Psi^{\prime} (D^M \Psi)^{*\prime} \right)\bigg|_{r=\infty_2}^{r=\infty_1} \nonumber \\
& + \int d^4x \int_{\infty_2}^{\infty_1} dr \sqrt{-g} \frac{1}{2}A_N^\prime \nabla_M F^{\prime MN} \nonumber \\
= & -\int d^4x \sqrt{-\gamma} n_M \left( \frac{1}{2} A_N^\prime F^{MN} + \frac{1}{2} \Psi^* D^M \Psi + \frac{1}{2} \Psi (D^M \Psi)^* \right)\bigg|_{r=\infty_2}^{r=\infty_1} \nonumber \\
& + \int d^4x \int_{\infty_2}^{\infty_1} dr \sqrt{-g} \frac{1}{2}A_N^\prime \nabla_M F^{MN}, \label{S0_prime}
\end{align}
where we have imposed the scalar's EOM and utilized gauge invariant property of some terms \eqref{gauge_invariant}. Via the gauge transformation \eqref{gauge_trans}, \eqref{S0_prime} becomes
\begin{align}
S_0 = & -\int d^4x \left\{ \sqrt{-g} \frac{1}{2} (A_\mu - \partial_\mu \Lambda) F^{r\mu} + \frac{1}{2}\sqrt{-\gamma} n_M \left[ \Psi^* D^M \Psi + \Psi (D^M \Psi)^* \right] \right\}\bigg|_{r=\infty_2}^{r=\infty_1} \nonumber \\
& + \int d^4x \int_{\infty_2}^{\infty_1} dr \left\{ \sqrt{-g} \frac{1}{2}A_N \nabla_M F^{MN} - \frac{1}{2} (\partial_N \Lambda) \sqrt{-g} \nabla_M F^{MN} \right\} \nonumber \\
= & -\int d^4x \left\{ \sqrt{-g} \frac{1}{2} (A_\mu - \partial_\mu \Lambda) F^{r\mu} + \frac{1}{2}\sqrt{-\gamma} n_M \left[ \Psi^* D^M \Psi + \Psi (D^M \Psi)^* \right] \right\}\bigg|_{r=\infty_2}^{r=\infty_1} \nonumber \\
& + \int d^4x \int_{\infty_2}^{\infty_1} dr \left\{ \sqrt{-g} \frac{1}{2}A_N \nabla_M F^{MN} - \frac{1}{2} \partial_N [\Lambda \sqrt{-g} \nabla_M F^{MN}] \right\} \nonumber \\
= & -\int d^4x \left\{ \sqrt{-g} \frac{1}{2} (A_\mu - \partial_\mu \Lambda) F^{r\mu} + \frac{1}{2}\sqrt{-\gamma} n_M \left[ \Psi^* D^M \Psi + \Psi (D^M \Psi)^* \right] \right\}\bigg|_{r=\infty_2}^{r=\infty_1} \nonumber \\
& + \int d^4x \int_{\infty_2}^{\infty_1} dr \sqrt{-g} \frac{1}{2}A_N \nabla_M F^{MN} - \frac{1}{2} \int d^4x \sqrt{-\gamma} n_N \Lambda \nabla_M F^{MN}  \nonumber \\
= & -\int d^4x \left\{ \sqrt{-g} \frac{1}{2} (A_\mu - \partial_\mu \Lambda) F^{r\mu} + \frac{1}{2}\sqrt{-\gamma} n_M \left[ \Psi^* D^M \Psi + \Psi (D^M \Psi)^* \right] \right\}\bigg|_{r=\infty_2}^{r=\infty_1} \nonumber \\
& + \int d^4x \int_{\infty_2}^{\infty_1} dr \sqrt{-g} \frac{1}{2}A_N \nabla_M F^{MN} + \frac{1}{2} \int d^4x \sqrt{-\gamma} n_N \partial_\mu \Lambda F^{\mu N} \nonumber \\
= & -\int d^4x \left\{ \sqrt{-g} \frac{1}{2} A_\mu F^{r\mu} + \frac{1}{2}\sqrt{-\gamma} n_M \left[ \Psi^* D^M \Psi + \Psi (D^M \Psi)^* \right] \right\}\bigg|_{r=\infty_2}^{r=\infty_1} \nonumber \\
& + \int d^4x \int_{\infty_2}^{\infty_1} dr \sqrt{-g} \frac{1}{2}A_N \nabla_M F^{MN}, \nonumber \\
= & -\int d^4x \sqrt{-\gamma} n_M \left\{ \frac{1}{2} A_N F^{MN} + \frac{1}{2}\Psi^* D^M \Psi + \frac{1}{2}\Psi (D^M \Psi)^* \right\}\bigg|_{r=\infty_2}^{r=\infty_1} \nonumber \\
& + \int d^4x \int_{\infty_2}^{\infty_1} dr \sqrt{-g} \frac{1}{2}A_N \nabla_M F^{MN},  \label{S0_any_gauge}
\end{align}
where in the second equality we have utilized the Bianchi identity
\begin{align}
\partial_N (\sqrt{-g} \nabla_M F^{MN})=0.
\end{align}
Moreover, in obtaining \eqref{S0_any_gauge} we have dropped some boundary derivative terms. So far, the gauge transformation \eqref{gauge_trans} is arbitrary. Thus, the result \eqref{S0_any_gauge} is valid for any gauge-fixing. The last line of \eqref{S0_any_gauge} can be further simplified by imposing dynamical EOMs. We will be less general and compare between two different gauge choices: $A_r =0 $ versus $A_r =- A_v/(r^2f(r))$.

First, we consider the radial gauge $A_r=0$, and the gauge transformation parameter is (up to a residual gauge):
\begin{align}
A_r =0 \Rightarrow \Lambda(r,x^\alpha) = \int_r^{r_c} dy A_r^{\prime}(y,x^\alpha).
\end{align}
With the set of dynamical EOMs \eqref{eom_dynamic_EF_alternative} imposed, \eqref{S0_any_gauge} turns into
\begin{align}
S_0 = & - \int d^4x \sqrt{-\gamma} n_M \left( \frac{1}{2} A_N F^{MN} + \frac{1}{2}\Psi^* D^M \Psi + \frac{1}{2}\Psi (D^M \Psi)^* \right)\bigg|_{r=\infty_2}^{r=\infty_1} \nonumber \\
&+ \int d^4x \int_{\infty_2}^{\infty_1} dr \sqrt{-g} \frac{1}{2}iq A_\mu [\Psi^* (D^\mu \Psi) - \Psi (D^\mu \Psi)^*] \nonumber \\
= & - \int d^4x \sqrt{-\gamma} n_M \left( \frac{1}{2} A_N F^{MN} + \frac{1}{2}\Psi^* D^M \Psi + \frac{1}{2}\Psi (D^M \Psi)^* \right)\bigg|_{r=\infty_2}^{r=\infty_1} \nonumber \\
&+ \int d^4x \int_{\infty_2}^{\infty_1} dr \sqrt{-g} \frac{1}{2}iq A_N [\Psi^* (D^N \Psi) - \Psi (D^N \Psi)^*].  \label{S0_pos_EF_radial}
\end{align}

Lets turn to a different gauge choice $A_r =- A_v/(r^2f(r))$, for which the gauge transformation parameter obeys
\begin{align}
A_r =- \frac{A_v}{r^2f(r)} \Rightarrow \left[ \partial_r + \frac{\partial_v}{ r^2f(r)}  \right] \Lambda = -A_r^\prime - \frac{A_v^\prime}{r^2f(r)}.
\end{align}
Accordingly, we will impose the set of dynamical EOMs \eqref{eom_dynamic_EF}. Then, \eqref{S0_any_gauge} becomes
\begin{align}
S_0 = & - \int d^4x \sqrt{-\gamma} n_M \left( \frac{1}{2} A_N F^{MN} + \frac{1}{2}\Psi^* D^M \Psi + \frac{1}{2}\Psi (D^M \Psi)^* \right)\bigg|_{r=\infty_2}^{r=\infty_1} \nonumber \\
&+ \int d^4x \int_{\infty_2}^{\infty_1} dr \sqrt{-g} \frac{1}{2}iq A_N [\Psi^* (D^N \Psi) - \Psi (D^N \Psi)^*], \label{S0_pos_Schw_radial1}
\end{align}
which takes the same form as that of \eqref{S0_pos_EF_radial}.

This confirms that the result \eqref{S0_pos_Schw_radial} is independent of gauge choice and thus the off-shell procedure is free of ambiguity, as long as dynamical EOMs are correctly taken according to gauge choice. With the help of \eqref{gauge_trans}, both \eqref{S0_pos_EF_radial} and \eqref{S0_pos_Schw_radial1} will be computed based on the gauge-fixed bulk solutions. Recall that the gauge-fixed solutions $(A_\mu, \Psi, \Psi^*)$ are functionals of $(\mathcal A_\mu^\prime + \partial_\mu \varphi, \Delta^\prime e^{iq \varphi}, \bar \Delta^\prime e^{-iq \varphi})$, so is $S_{eff}$:
\begin{align}
S_{eff}= S_{eff}[\mathcal A_\mu^\prime + \partial_\mu \varphi, \Delta^\prime e^{iq \varphi}, \bar \Delta^\prime e^{-iq \varphi}].
\end{align}

To summarize, in the bulk the low energy hydrodynamic field is identified with the gauge transformation parameter bringing the field configuration without any gauge-fixing into a specific gauge. Using the solution obtained in this specific gauge-fixing, we implement the bulk path integral (in the partially on-shell sense) and eventually identify the renormalized partially on-shell bulk action as the boundary effective action.

\section{Source terms in the perturbative EOMs} \label{sources}

In this appendix, we record explicit formulas for various source terms in the perturbative EOMs \eqref{eom_Box}.
%
%Since we wish to obtain the boundary effective action $S_{eff}$ up to the order $\mathcal O(\xi^1 \lambda^2 \alpha^1)$, we will truncate the triple expansion of bulk fields. Specifically, we will solve $A_v$ up to $\mathcal O(\xi^1 \lambda^2 \alpha^1)$, and $\Psi, \Psi^*$ up to $\mathcal O(\xi^1 \lambda^1 \alpha^1)$.
%
First, we consider the zeroth order in time-derivative expansion. At the order $\mathcal{O}(\xi^0 \lambda^1 \alpha^1)$, we have
\begin{align}
j_\Psi^{(0)(1)(1)}= & - \frac{2q^2r}{f(r)} \bar A_v \delta \bar A_v \Psi^{(0)(1)(0)}, \nonumber \\
j_{\Psi^*}^{(0)(1)(1)} = & - \frac{2q^2r}{f(r)} \bar A_v \delta \bar A_v \Psi^{*(0)(1)(0)}. \label{jPsi_011}
\end{align}
At the next orders $\mathcal{O}(\xi^0 \lambda^2 \alpha^0)$ and $\mathcal{O}(\xi^0 \lambda^2 \alpha^1)$:
\begin{align}
j_v^{(0)(2)(0)} = &\frac{2q^2r}{f(r)} \bar A_v \Psi^{(0)(1)(0)} \Psi^{*(0)(1)(0)}, \nonumber \\
j_v^{(0)(2)(1)} %= & \frac{2q^2r}{f(r)} \bar A_v \left[\Psi^{(0)(1)(1)} \Psi^{*(0)(1)(0)} + \Psi^{(0)(1)(0)} \Psi^{*(0)(1)(1)} \right] \nonumber \\
%& + \frac{2q^2r}{f(r)} \delta\bar A_v \Psi^{(0)(1)(0)} \Psi^{*(0)(1)(0)}, \nonumber \\
= & \frac{2q^2r}{f(r)} \bar A_v \left[\Psi^{(0)(1)(1)} \Psi^{*(0)(1)(0)} + \Psi^{(0)(1)(0)} \Psi^{*(0)(1)(1)} \right] \nonumber \\
& + \mu_0^{-1} \delta \mu j_v^{(0)(2)(0)}.
\label{jv_020_021}
\end{align}
We turn to time-derivative corrections. First of all, as the chemical potential $\mu$ is taken as a constant, $A_v^{(1)(0)(0)}= A_v^{(1)(0)(1)}=0$, which simplifies subsequent calculations. At the order $\mathcal{O}(\xi^1 \lambda^1 \alpha^0)$, we have
\begin{align}
j_\Psi^{(1)(1)(0)} = & -2r^3 \partial_r \partial_v \Psi^{(0)(1)(0)} - 3r^2 \partial_v \Psi^{(0)(1)(0)} - \frac{2iq r}{f(r)} \bar A_v \partial_v \Psi^{(0)(1)(0)}, \nonumber \\
j_{\Psi^*}^{(1)(1)(0)} = & -2r^3 \partial_r \partial_v \Psi^{*(0)(1)(0)} - 3r^2 \partial_v \Psi^{*(0)(1)(0)} + \frac{2iq r}{f(r)} \bar A_v \partial_v \Psi^{*(0)(1)(0)}. \label{jPsi_110}
\end{align}
At the order $\mathcal{O}(\xi^1 \lambda^2 \alpha^0)$,
\begin{align}
j_v^{(1)(2)(0)} = & - \left[ \frac{2r}{f(r)}\partial_r + \frac{1}{f(r)} - \frac{rf^\prime(r)}{f^2(r)} \right] \partial_v A_v^{(0)(2)(0)} \nonumber \\
&+ \frac{iqr}{f(r)} \left( \Psi^{*(0)(1)(0)} \partial_v \Psi^{(0)(1)(0)} - \Psi^{(0)(1)(0)} \partial_v \Psi^{*(0)(1)(0)}\right) \nonumber \\
& + \frac{2q^2r}{f(r)} \bar A_v\left( \Psi^{*(1)(1)(0)} \Psi^{(0)(1)(0)} + \Psi^{*(0)(1)(0)} \Psi^{(1)(1)(0)} \right). \label{jv_120}
\end{align}

In order to see quartic terms in $S_{eff}$, it will be sufficient to stick to the static limit and critical point, namely, we just need to solve $\Psi^{(0)(3)(0)}$, $\Psi^{*(0)(3)(0)}$, and $A_v^{(0)(4)(0)}$. The reason is that quartic terms do not vanish at the critical value $\mu_0 = 2r_h^2$, as indicated by the field theory result \cite{Levchenko2007,Kamenev2011}. Then, if we want to compute $S_{eff}$ at the order $\mathcal{O}(\xi^0 \lambda^4 \alpha^0)$, we need the following source terms
\begin{align}
j_\Psi^{(0)(3)(0)} = & - \frac{2q^2r}{f(r)} \bar A_v A_v^{(0)(2)(0)} \Psi^{(0)(1)(0)}, \nonumber \\
j_{\Psi^*}^{(0)(3)(0)} = & - \frac{2q^2r}{f(r)} \bar A_v A_v^{(0)(2)(0)} \Psi^{*(0)(1)(0)}, \label{jPsi_030}
\end{align}
and
\begin{align}
j_v^{(0)(4)(0)} = &\frac{2q^2r}{f(r)} \left[ \Psi^{*(0)(1)(0)} \Psi^{(0)(1)(0)} A_v^{(0)(2)(0)} + \Psi^{*(0)(3)(0)} \Psi^{(0)(1)(0)} \bar A_v \right. \nonumber \\
& \qquad  \left. + \Psi^{*(0)(1)(0)} \Psi^{(0)(3)(0)} \bar A_v \right]. \label{jv_040}
\end{align}

\section*{Acknowledgements}

MF would like to thank Masaki Tezuka for helpful discussions and comments. YB was supported by the Natural Science Foundation of China (NSFC) under the grant No. 11705037. SL was supported by NSFC under Grant Nos 11675274 and 11735007.

\bibliographystyle{utphys}
\bibliography{reference}
\end{document}